# Generic energy formalism for reciprocal quadruplets within the two-sublattice quasichemical model

Kun Wang[1*], Patrice Chartrand[2]

*1 State Key Laboratory of Advanced Special Steel & Shanghai Key Laboratory of Advanced Ferrometallurgy & School of Materials Science and Engineering, Shanghai University, Shanghai, 200072, China*

*2 Center for Research in Computational Thermochemistry (CRCT), Department of Chemical Engineering, Polytechnique Montréal, Montréal H3C 3A7, Québec, Canada*

Abstract: Ever-increasing interests for more accurate thermodynamic predictions of phase diagrams motivate the development of more reliable thermodynamic models. The Modified Quasichemical Model within the two-sublattice Quadruplet Approximation (MQMQA) was thus established in well response to these interests and motivations. However, the model still needs to be further improved in order to have better thermodynamic predictions of general reciprocal solutions. The present paper proposes a new and generic formalism to characterize the Gibbs energy of the ternary reciprocal quadruplet within the framework of the MQMQA. The new formalism is developed by circumventing the problem spawned from solving the singular matrix of mass equations. As a result, energy landscapes of reciprocal solutions can be better defined everywhere in reciprocal composition spaces. With the current improvement, the MQMQA is believed to be one of the most reliable thermodynamic models for various types of solutions with or without short-range ordering.

## 1. Introduction

Thermodynamics and phase equilibria are essential to materials science and technology [1]. To date, the Calphad method is still the sole way to calculate them in multicomponent systems. The simulation accuracy is largely dependent on whether the selected models can precisely describe the thermodynamic properties of various phases in materials regardless of their measured data zones or unexplored regions. Models based on the Bragg-Williams or mean-field approximations [2-3] are mathematical representatives with ideal entropy of mixing. They are not suitable to treat a solution with Short-Range Ordering (SRO) where the entropy of mixing as

E-mail address: wangkun0808@shu.edu.cn

a function of composition takes on an "m-shaped" curve while the enthalpy of mixing displays a "V-shaped" curve [4] as shown in Fig.1. The associated solution model [5-7] could partially address these issues involved in the solution with SRO; however, the entropy paradox is not avoidable where the entropy of mixing will never return to be ideal when no interactions are appeared between components. These problems are common and well known in mathematical models. They have made adverse impacts on the predictive power towards those unknown regions in multicomponent systems.

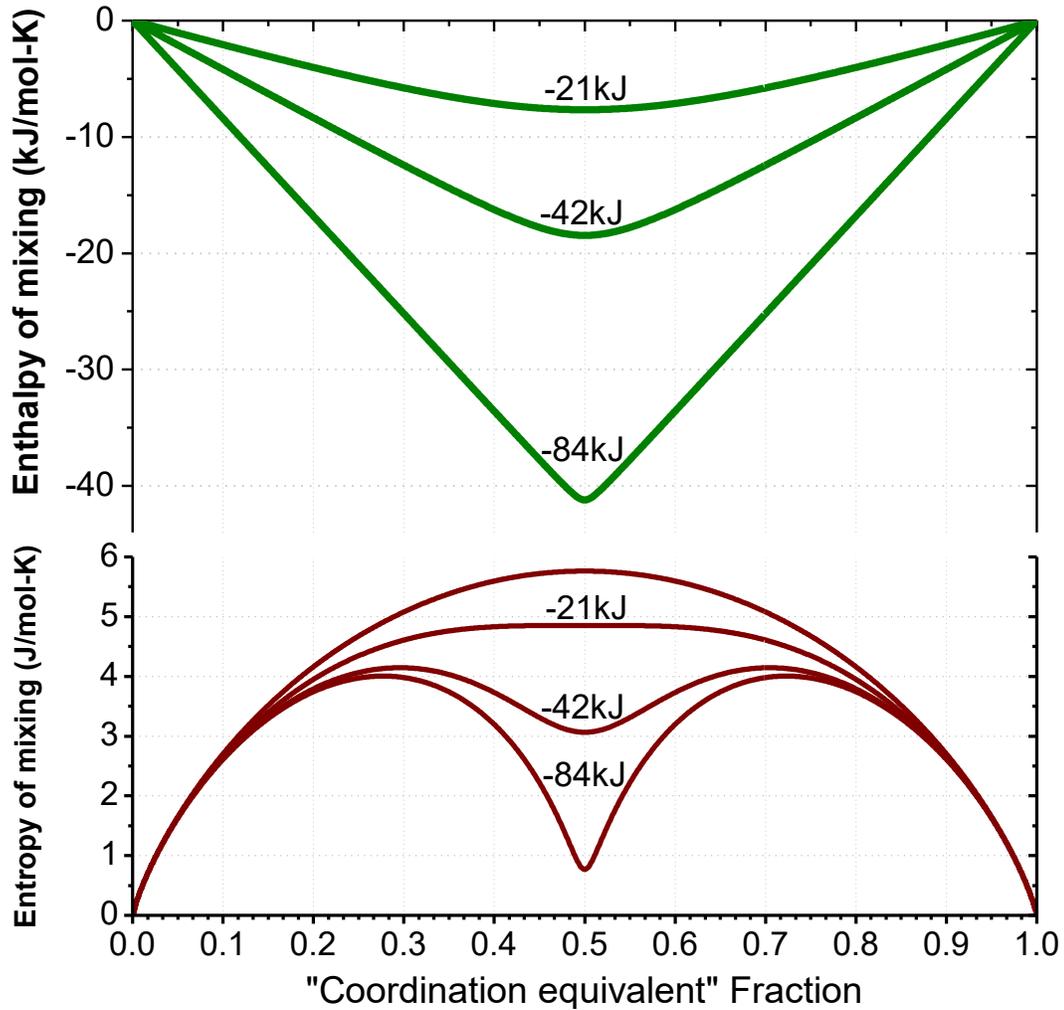

Fig. 1 Molar enthalpy and entropy of mixing for an arbitrary A-B system calculated at 1000 °C from the quasichemical model for short-range ordering with $\Delta G_{AB} = 0, -21, -42,$ and -84 kJ/mol and with $Z_A=Z_B=2$ [8]

In order to enable reliable predictions, A. D. Pelton and his colleagues spent several decades developing a thermodynamic model dependent on the solution structure [8-11]. It was based upon the classical quasichemical model of Fowler and Guggenheim [4]. Several important modifications were made to the model of Fowler and Guggenheim: (1) permit the composition of maximum SRO in a binary system to be freely chosen, (2) allow the coordination numbers to vary with composition, (3) express the formation energy of structural entities (pairs or quadruplets) as a polynomial in the entity fractions, and (4) to extend the model to multicomponent systems. The model has recently been used to critically evaluate and optimize several hundred liquid oxide [12-15], salt [16-23], and alloy solutions [24-26]. Afterwards, the model was further extended to two-sublattice liquids [11]. This is the so-called MQMQA with particular renown in the application to reciprocal liquids having strong SROs from both the First-Nearest Neighbor (FNN) and Second-Nearest Neighbor (SNN) interactions. In view of the fruitful MQMQA, McMurray and Besmann et al [27] are employing the MQMQA to treat molten salts (common-ion and reciprocal) and expanding the molten salt database for molten salt nuclear reactors. However, the MQMQA with the two-sublattice formalism is very complex, and furthermore, the quasichemical reaction equation between a reciprocal quadruplet and all its surrounding binary quadruplets was not strictly balanced (seeing Eqs.37-38 in [11]). The balancing issue may adversely influence, from practical aspects, the accurate definition of the default Gibbs energy for the reciprocal quadruplet, and from theoretical aspects, the theoretical rigor of the model. In extreme cases, the balancing issue may cause artificial phase relations to be predicted, weakening the predictive ability that the structural MQMQA should intrinsically possesses.

The present paper proposes an innovative way to rigorously balance the quasichemical reaction equation between the reciprocal quadruplet and all its surrounding binary quadruplets. The model equation describing the energy of the reciprocal quadruplet is thus established in a consistent and symmetrical fashion by employing the energies of all the constituent binary quadruplets. No matter what the chemistry of the reciprocal SRO is, its default Gibbs energy can be correctly defined without the balancing issue. Furthermore, the main equations of the MQMQA are reformulated in a logical, straightforward and readable fashion. All these improvements are designed to render the MQMQA a more reliable thermodynamic model for various types of solutions with or without SRO.

## 2. Elucidation of MQMQA

The MQMQA [11] is a two-sublattice model developed to treat various SROs (FNN, SNN and reciprocal) in reciprocal solutions by means of quadruplets. One quadruplet includes two cations and two anions forming four FNN pairs and two SNN pairs. Correspondingly, a total of nine quadruplets can be generated in a ternary reciprocal solution. The relative energies of various quadruplets (unary, binary and reciprocal) dominate their relative amounts in the solution and can thus simultaneously describe various SROs in particular systems [19, 23]. The model can also be extended to multicomponent systems with more cations and anions by proper geometrical interpolations [28-29]. For simplicity, the A, B//X, Y prototype is adopted to describe the model details. However, the newly developed formalism based on this prototype will never influence its application to multicomponent reciprocal systems. The Readers are referred to the literature [10] for details regarding the multicomponent MQMQA.

2.1 The MQMQA formalism

The MQMQA formalism for an A, B//X, Y solution can be expressed as follows,

$$G = n_{A_2X_2}G_{A_2X_2} + n_{B_2X_2}G_{B_2X_2} + n_{A_2Y_2}G_{A_2Y_2} + n_{B_2Y_2}G_{B_2Y_2} + n_{ABX_2}G_{ABX_2} + n_{ABY_2}G_{ABY_2}$$
$$+ n_{A_2XY}G_{A_2XY} + n_{B_2XY}G_{B_2XY} + n_{ABXY}G_{ABXY} - T\Delta S^{config} \quad (1)$$

where $n_{A_2X_2}$ and $G_{A_2X_2}$ are the number of moles and the molar Gibbs free energy of the $A_2X_2$ quadruplet ([$A_2X_2$]$_{quad}$, similarly for the others), respectively, and T is temperature in Kelvins. $\Delta S^{config}$ is the configurational entropy of mixing of various quadruplets on the quadruplet lattice sites; the mathematical formulae will be illustrated in the following section. As shown in Fig. 2, there are nine quadruplets in the A, B//X, Y compositional space. Each quadruplet consists of two cations and two anions, thus resulting in four FNN and two SNN pairs. For example, the reciprocal [ABXY]$_{quad}$ contains A-X, B-X, A-Y and B-Y as the FNN pairs, and A–B and X–Y as the SNN pairs. Actually, the Gibbs energy of such a solution originates from the summation over all the quadruplets' energies along with the configurational entropy. The nine quadruplets in the A, B//X, Y solution are grouped as four unary quadruplets with apparent chemical formulas as [$A_2X_2$]$_{quad}$, [$B_2X_2$]$_{quad}$, [$A_2Y_2$]$_{quad}$ and [$B_2Y_2$]$_{quad}$, and four binary quadruplets as [ABX$_2$]$_{quad}$, [ABY$_2$]$_{quad}$, [$A_2$XY]$_{quad}$ and [$B_2$XY]$_{quad}$, and one ternary reciprocal quadruplet as [ABXY]$_{quad}$. The actual chemistries of these quadruplets are determined by the SNN coordination numbers of ions on each quadruplet as shown in Table 1. $Z^A_{ABXY}$ represents the SNN coordination number of ion A in the ABXY quadruplet. The real chemistry of [ABXY]$_{quad}$ is $A_{\frac{1}{Z^A_{ABXY}}} B_{\frac{1}{Z^B_{ABXY}}} X_{\frac{1}{Z^X_{ABXY}}} Y_{\frac{1}{Z^Y_{ABXY}}}$, and there are similar formulas for all the other quadruplets. Each quadruplet must be a charge-neutrality entity, which leads to the following equation:

$$\frac{q_A}{Z^A_{ABXY}} + \frac{q_B}{Z^B_{ABXY}} = \frac{q_X}{Z^X_{ABXY}} + \frac{q_Y}{Z^Y_{ABXY}} \tag{2}$$

where $q_A$, $q_B$, $q_X$ and $q_Y$ are the absolute ionic charges of A, B, X and Y, respectively. This equation also holds when A=B and/or when X=Y, as listed in Table 1. For unary quadruplets, their Gibbs energies are directly extracted from those of the corresponding unary components considering the mass balance. For example, an [Al$_2$O$_2$]$_{quad}$ has the real chemistry as $Al_{\frac{2}{6}} O_{\frac{2}{4}}$, and its molar Gibbs energy is equal to 1/6 $G^0_{Al_2O_3}$. For a binary [ABX$_2$]$_{quad}$, its Gibbs energy is taken from the two constituent unary [A$_2$X$_2$]$_{quad}$ and [B$_2$X$_2$]$_{quad}$ as:

$$[A_2X_2]_{quad} + [B_2X_2]_{quad} = 2[ABX_2]_{quad} \quad 2\Delta G_{ABX_2} \tag{3}$$

where $\Delta G_{ABX_2}$ is the Gibbs energy of formation of one mole of [ABX$_2$]$_{quad}$. As indicated above, the real chemistries of [A$_2$X$_2$]$_{quad}$, [B$_2$X$_2$]$_{quad}$ and [ABX$_2$]$_{quad}$ are $A_{\frac{2}{Z^A_{A_2X_2}}} X_{\frac{2}{Z^X_{A_2X_2}}}$, $B_{\frac{2}{Z^B_{B_2X_2}}} X_{\frac{2}{Z^X_{B_2X_2}}}$ and $A_{\frac{1}{Z^A_{ABX_2}}} B_{\frac{1}{Z^B_{ABX_2}}} X_{\frac{2}{Z^X_{ABX_2}}}$, respectively. Hence, the schematic reaction of equation (3) must be altered to the following reaction as,

$$m\, A_{\frac{2}{Z^A_{A_2X_2}}} X_{\frac{2}{Z^X_{A_2X_2}}} + n\, B_{\frac{2}{Z^B_{B_2X_2}}} X_{\frac{2}{Z^X_{B_2X_2}}} = 2 A_{\frac{1}{Z^A_{ABX_2}}} B_{\frac{1}{Z^B_{ABX_2}}} X_{\frac{2}{Z^X_{ABX_2}}} \tag{4}$$

where the balancing coefficients $m$ and $n$ are solved according to the mass conservation between both sides of the equation. With the mass conservation of A, $m$ is derived to be $\frac{Z^A_{A_2X_2}}{Z^A_{ABX_2}}$. Following the same procedure for B, $n$ is determined to be $\frac{Z^B_{B_2X_2}}{Z^B_{ABX_2}}$. Based on the charge-neutrality condition, the mass balance of X is inherently maintained. Thus, the Gibbs energy of one mole of [ABX$_2$]$_{quad}$ is deduced as,

$$G_{ABX_2} = \frac{Z^A_{A_2X_2}}{2Z^A_{ABX_2}} G^0_{A_2X_2} + \frac{Z^B_{B_2X_2}}{2Z^B_{ABX_2}} G^0_{B_2X_2} + \Delta G_{ABX_2} \tag{5}$$

where $\Delta G_{ABX_2}$ is a polynomial function of quadruplet fractions. It is formulated as,

$$\Delta G_{ABX_2} = \Delta G^0_{ABX_2} + (\Delta G_{ABX_2} - \Delta G^0_{ABX_2}) \tag{6}$$

where $\Delta G^0_{ABX_2}$ is a constant at a fixed temperature, independent of composition, and $(\Delta G_{ABX_2} - \Delta G^0_{ABX_2})$ is expanded as an empirical polynomial in the quadruplet fractions $X_{A_2X_2}$ and/or $X_{B_2X_2}$ (similarly for others). The quadruplet fractions will be defined and interpreted in section 3 and the Appendix. Here, the standard molar Gibbs energy of $[ABX_2]_{quad}$ is introduced as,

$$G^0_{ABX_2} = \frac{Z^A_{A_2X_2}}{2Z^A_{ABX_2}} G^0_{A_2X_2} + \frac{Z^B_{B_2X_2}}{2Z^B_{ABX_2}} G^0_{B_2X_2} + \Delta G^0_{ABX_2} \tag{7}$$

and equation (5) can thus be rearranged as,

$$G_{ABX_2} = G^0_{ABX_2} + (\Delta G_{ABX_2} - \Delta G^0_{ABX_2}) \tag{8}$$

A similar path can be taken for defining the Gibbs energies of all the other binary quadruplets. For the reciprocal $[ABXY]_{quad}$, its Gibbs energy should also be drawn from those of the corresponding binary quadruplets. The schematic reaction proposed by Pelton [11] is shown as,

$$\frac{1}{2}\left([ABX_2]_{quad} + [ABY_2]_{quad} + [A_2XY]_{quad} + [B_2XY]_{quad}\right) = 2[ABXY]_{quad} \quad 2\Delta G_{ABXY} \tag{9}$$

where $\Delta G_{ABXY}$ is the Gibbs energy of formation of one mole of $[ABXY]_{quad}$ from all the constituent binary quadruplets. The schematic reaction (9) has also to be transformed to the real reaction as,

$$\frac{1}{2}\left(m\, A_{\frac{1}{Z^A_{ABX_2}}} B_{\frac{1}{Z^B_{ABX_2}}} X_{\frac{2}{Z^X_{ABX_2}}} + n\, B_{\frac{2}{Z^B_{B_2XY}}} X_{\frac{1}{Z^X_{B_2XY}}} Y_{\frac{1}{Z^Y_{B_2XY}}} + \bar{o}\, A_{\frac{1}{Z^A_{ABY_2}}} B_{\frac{1}{Z^B_{ABY_2}}} Y_{\frac{2}{Z^Y_{ABY_2}}} + p\, A_{\frac{2}{Z^A_{A_2XY}}} X_{\frac{1}{Z^X_{A_2XY}}} Y_{\frac{1}{Z^Y_{A_2XY}}}\right) = 2 A_{\frac{1}{Z^A_{ABXY}}} B_{\frac{1}{Z^B_{ABXY}}} X_{\frac{1}{Z^X_{ABXY}}} Y_{\frac{1}{Z^Y_{ABXY}}} \tag{10}$$

where the balancing coefficients $m$, $n$, $\bar{o}$ and $p$ need to be determined by the mass balance for all ions. The mass-balance equations are displayed as follows,

$$\frac{m}{Z^A_{ABX_2}} + \frac{\bar{o}}{Z^A_{ABY_2}} + \frac{2p}{Z^A_{A_2XY}} = \frac{4}{Z^A_{ABXY}} \tag{11}$$

$$\frac{m}{Z^B_{ABX_2}} + \frac{2n}{Z^B_{B_2XY}} + \frac{\bar{o}}{Z^B_{ABY_2}} = \frac{4}{Z^B_{ABXY}} \tag{12}$$

$$\frac{2m}{Z^X_{ABX_2}} + \frac{n}{Z^X_{B_2XY}} + \frac{p}{Z^X_{A_2XY}} = \frac{4}{Z^X_{ABXY}} \tag{13}$$

$$\frac{n}{Z^Y_{B_2XY}} + \frac{2\bar{o}}{Z^Y_{ABY_2}} + \frac{p}{Z^Y_{A_2XY}} = \frac{4}{Z^Y_{ABXY}} \tag{14}$$

These linear equations can be rewritten in matrix form as,

$$\overbrace{\begin{bmatrix} \frac{1}{Z^A_{ABX_2}} & 0 & \frac{1}{Z^A_{ABY_2}} & \frac{2}{Z^A_{A_2XY}} \\ \frac{1}{Z^B_{ABX_2}} & \frac{2}{Z^B_{B_2XY}} & \frac{1}{Z^B_{ABY_2}} & 0 \\ \frac{2}{Z^X_{ABX_2}} & \frac{1}{Z^X_{B_2XY}} & 0 & \frac{1}{Z^X_{A_2XY}} \\ 0 & \frac{1}{Z^Y_{B_2XY}} & \frac{2}{Z^Y_{ABY_2}} & \frac{1}{Z^Y_{A_2XY}} \end{bmatrix}}^{A} \overbrace{\begin{bmatrix} m \\ n \\ \bar{o} \\ p \end{bmatrix}}^{x} = \overbrace{\begin{bmatrix} \frac{4}{Z^A_{ABXY}} \\ \frac{4}{Z^B_{ABXY}} \\ \frac{4}{Z^X_{ABXY}} \\ \frac{4}{Z^Y_{ABXY}} \end{bmatrix}}^{B} \quad (15)$$

Unfortunately, the coefficient vector x cannot be solved since the rank of matrix $A$ is equal to 3, which means there are only three independent equations but with four unknown coefficients to be defined. The implied mechanism is due to the charge-neutrality condition of equation (2), which decreases the matrix rank. Pelton [11] has proposed the following equation as,

$$G_{ABXY} = (\frac{q_X}{Z^X_{ABXY}} + \frac{q_Y}{Z^Y_{ABXY}})^{-1} \left( \frac{q_X Z^A_{A_2X_2}}{2Z^A_{ABXY} Z^X_{ABXY}} G_{A_2X_2} + \frac{q_X Z^B_{B_2X_2}}{2Z^B_{ABXY} Z^X_{ABXY}} G_{B_2X_2} + \frac{q_Y Z^A_{A_2Y_2}}{2Z^A_{ABXY} Z^Y_{ABXY}} G_{A_2Y_2} + \frac{q_Y Z^B_{B_2Y_2}}{2Z^B_{ABXY} Z^Y_{ABXY}} G_{B_2Y_2} \right) +$$
$$\frac{1}{4}(m\Delta G_{ABX_2} + n\Delta G_{B_2XY} + \bar{o}\Delta G_{ABY_2} + p\Delta G_{A_2XY}) + \Delta G_{ABXY} \quad (16)$$

where $m = \frac{Z^X_{ABX_2}}{Z^X_{ABXY}}, n = \frac{Z^B_{B_2XY}}{Z^B_{ABXY}}, \bar{o} = \frac{Z^Y_{ABY_2}}{Z^Y_{ABXY}}$ and $p = \frac{Z^A_{A_2XY}}{Z^A_{ABXY}}$ were defined. However, this is strictly correct only if the following conditions are met,

$$\frac{\frac{q_A}{Z^A_{ABXY}}}{\frac{q_A}{Z^A_{ABXY}} + \frac{q_B}{Z^B_{ABXY}}} = \frac{\frac{q_A}{Z^A_{ABX_2}}}{\frac{q_A}{Z^A_{ABX_2}} + \frac{q_B}{Z^B_{ABX_2}}} = \frac{\frac{q_A}{Z^A_{ABY_2}}}{\frac{q_A}{Z^A_{ABY_2}} + \frac{q_B}{Z^B_{ABY_2}}} = Y'_A \quad (17)$$

$$\frac{\frac{q_B}{Z^B_{ABXY}}}{\frac{q_A}{Z^A_{ABXY}} + \frac{q_B}{Z^B_{ABXY}}} = \frac{\frac{q_B}{Z^B_{ABX_2}}}{\frac{q_A}{Z^A_{ABX_2}} + \frac{q_B}{Z^B_{ABX_2}}} = \frac{\frac{q_B}{Z^B_{ABY_2}}}{\frac{q_A}{Z^A_{ABY_2}} + \frac{q_B}{Z^B_{ABY_2}}} = Y'_B \quad (18)$$

$$\frac{\frac{q_X}{Z^X_{ABXY}}}{\frac{q_X}{Z^X_{ABXY}} + \frac{q_Y}{Z^Y_{ABXY}}} = \frac{\frac{q_X}{Z^X_{A_2XY}}}{\frac{q_X}{Z^X_{A_2XY}} + \frac{q_Y}{Z^Y_{A_2XY}}} = \frac{\frac{q_X}{Z^X_{B_2XY}}}{\frac{q_X}{Z^X_{B_2XY}} + \frac{q_Y}{Z^Y_{B_2XY}}} = Y'_X \quad (19)$$

$$\frac{\frac{q_Y}{Z^Y_{ABXY}}}{\frac{q_X}{Z^X_{ABXY}} + \frac{q_Y}{Z^Y_{ABXY}}} = \frac{\frac{q_Y}{Z^Y_{A_2XY}}}{\frac{q_X}{Z^X_{A_2XY}} + \frac{q_Y}{Z^Y_{A_2XY}}} = \frac{\frac{q_Y}{Z^Y_{B_2XY}}}{\frac{q_X}{Z^X_{B_2XY}} + \frac{q_Y}{Z^Y_{B_2XY}}} = Y'_Y \quad (20)$$

where $Y'_A, Y'_B, Y'_X$ and $Y'_Y$ are the "charge-equivalent" fractions of ions A, B, X and Y, respectively. Let us firstly check the mass balance for the reciprocal [KBeFCl]$_{quad}$ shown below,

$$\frac{1}{2}(mK_{\frac{1}{z^K_{KBeF_2}}}Be_{\frac{1}{z^{Be}_{KBeF_2}}}F_{\frac{2}{z^F_{KBeF_2}}} + nBe_{\frac{2}{z^{Be}_{Be_2FCl}}}F_{\frac{1}{z^F_{Be_2FCl}}}Cl_{\frac{1}{z^{Cl}_{Be_2FCl}}} + \bar{o}K_{\frac{1}{z^K_{KBeCl_2}}}Be_{\frac{1}{z^{Be}_{KBeCl_2}}}Cl_{\frac{2}{z^{Cl}_{KBeCl_2}}} +$$
$$pK_{\frac{2}{z^K_{K_2FCl}}}F_{\frac{1}{z^F_{K_2FCl}}}Cl_{\frac{1}{z^{Cl}_{K_2FCl}}}) = 2K_{\frac{1}{z^K_{KBeFCl}}}Be_{\frac{1}{z^{Be}_{KBeFCl}}}F_{\frac{1}{z^F_{KBeFCl}}}Cl_{\frac{1}{z^{Cl}_{KBeFCl}}} \quad (21)$$

where the coordination numbers are taken from Wang et al. [23]. Using the definitions of balancing coefficients in equation (16), $m=3/3.2$, $n=4.8/6.4$, $\bar{o}=3/3.2$ and $p=6/3.2$ are obtained and then substituted into equation (21), causing the ratios of ions K, Be, F, Cl on the left side to be consistent with the chemistry of the reciprocal quadruplet on the right side. The reason is that all the coordination numbers were defined according to equations (17-20). If the real chemistry of the reciprocal quadruplet is $K_{\frac{1}{4}}Be_{\frac{1}{6}}F_{\frac{1}{3}}Cl_{\frac{1}{4}}$, and those of all the constituent binary quadruplets are $K_{\frac{1}{2}}Be_{\frac{1}{6}}F_{\frac{2}{2.4}}$, $K_{\frac{1}{6}}Be_{\frac{1}{3}}Cl_{\frac{2}{2.4}}$, $K_{\frac{2}{4}}F_{\frac{1}{3}}Cl_{\frac{1}{6}}$ and $Be_{\frac{2}{6}}F_{\frac{1}{2}}Cl_{\frac{1}{6}}$, then $m=2.4/3$, $n=6/6$, $\bar{o}=2.4/4$ and $p=4/4$ are substituted again into equation (21), and finally the ratios of ions K, Be, F, Cl are $\frac{1}{4}, \frac{1}{6}, \frac{3}{8}, \frac{5}{24}$ on the left side compared to $\frac{1}{4}, \frac{1}{6}, \frac{1}{3}, \frac{1}{4}$ on the right side, respectively. Although the formalism proposed by Pelton et al. [11] is very simple, it cannot completely balance the quasichemical reaction equation between the "reactants" and "products" when the interrelations of coordination numbers do not respect equations (17-20). Moreover, the above deficiency may result in a scenario where the calculated curve shape is inconsistent with what the rigorous theory intrinsically predicts. Let us randomly define a surrogate *ABXY* solution also having the reciprocal quadruplet $A_{\frac{1}{4}}B_{\frac{1}{6}}X_{\frac{1}{3}}Y_{\frac{1}{4}}$ and the constituent binary quadruplets $A_{\frac{1}{2}}B_{\frac{1}{6}}X_{\frac{2}{2.4}}$, $A_{\frac{1}{6}}B_{\frac{1}{3}}Y_{\frac{2}{2.4}}$, $A_{\frac{2}{4}}X_{\frac{1}{3}}Y_{\frac{1}{6}}$ and $B_{\frac{2}{6}}X_{\frac{1}{2}}Y_{\frac{1}{6}}$. For simplicity, we assume $\Delta G_{ABX_2} = \Delta G_{ABY_2} = \Delta G_{A_2XY} = \Delta G_{B_2XY} = -10$ kJ/mol and the energies of all the unary quadruplets to be nil. From equation (9), when $\Delta G_{ABXY} = 0$, all the involved quadruplets will be randomly distributed on the quadruplets' lattice sites, and the calculated energy minimum should be located at the composition of the reciprocal quadruplet. However, using the formalism proposed by Pelton et al. [11], the calculated energy minimum drifts a bit away from the expected place. This deviation can be seen from the blue curves in Fig. 3. The blue curves are the Gibbs energies of the reciprocal *ABXY* solution calculated by the previous formalism [11] along the $A_{\frac{1}{2}}B_{\frac{1}{4}}X_{\frac{2}{2}} - A_{\frac{1}{3}}B_{\frac{1}{3}}Y_{\frac{2}{2}}$ line (see red dashed line in Fig.2) where the reciprocal quadruplet $A_{\frac{1}{4}}B_{\frac{1}{6}}X_{\frac{1}{3}}Y_{\frac{1}{4}}$ is located. Decreasing $\Delta G_{ABXY}$ from 0 J/mol to -500 J/mol and -1000 J/mol, the extremum gradually approaches the composition of the reciprocal quadruplet due to the enhancement of its SRO. However, increasing the reciprocal SRO inevitably enlarges the calculated deviation from reality in energy using the previous formalism [11]. Hence, the above-mentioned issues may occur when the coordination numbers have not fulfilled the certain conditions (eqs.17-20). As a result, the previous formalism should be modified in order to present more realistic energy landscapes for general reciprocal solutions.

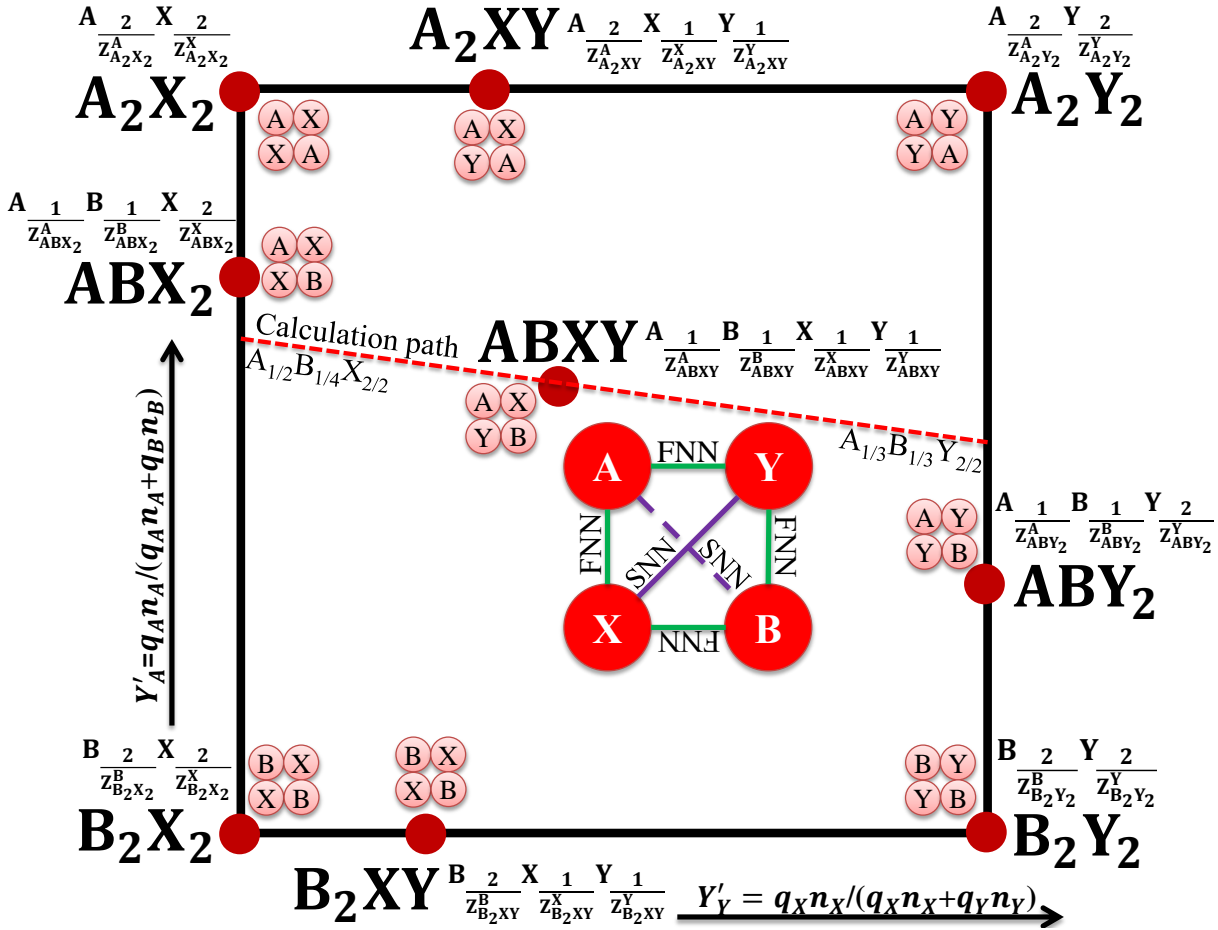

Fig. 2 Composition square of the AB//XY reciprocal ternary system showing various quadruplets

Table 1 Coordination numbers for ions in various quadruplets

| Quads \ Ions | 1st | 2nd | 3rd | 4th | Chemistry | Charge-neutrality condition |
|---|---|---|---|---|---|---|
| $[A_2X_2]_{quad}$ | $Z^A_{A_2X_2}$ | $Z^A_{A_2X_2}$ | $Z^X_{A_2X_2}$ | $Z^X_{A_2X_2}$ | $A_{\frac{2}{Z^A_{A_2X_2}}} X_{\frac{2}{Z^X_{A_2X_2}}}$ | $\frac{2q_A}{Z^A_{A_2X_2}} = \frac{2q_X}{Z^X_{A_2X_2}} = Q_{A_2X_2}$ |
| $[B_2X_2]_{quad}$ | $Z^B_{B_2X_2}$ | $Z^B_{B_2X_2}$ | $Z^X_{B_2X_2}$ | $Z^X_{B_2X_2}$ | $B_{\frac{2}{Z^B_{B_2X_2}}} X_{\frac{2}{Z^X_{B_2X_2}}}$ | $\frac{2q_B}{Z^B_{B_2X_2}} = \frac{2q_X}{Z^X_{B_2X_2}} = Q_{B_2X_2}$ |
| $[A_2Y_2]_{quad}$ | $Z^A_{A_2Y_2}$ | $Z^A_{A_2Y_2}$ | $Z^Y_{A_2Y_2}$ | $Z^Y_{A_2Y_2}$ | $A_{\frac{2}{Z^A_{A_2Y_2}}} Y_{\frac{2}{Z^Y_{A_2Y_2}}}$ | $\frac{2q_A}{Z^A_{A_2Y_2}} = \frac{2q_Y}{Z^Y_{A_2Y_2}} = Q_{A_2Y_2}$ |
| $[B_2Y_2]_{quad}$ | $Z^B_{B_2Y_2}$ | $Z^B_{B_2Y_2}$ | $Z^Y_{B_2Y_2}$ | $Z^Y_{B_2Y_2}$ | $B_{\frac{2}{Z^B_{B_2Y_2}}} Y_{\frac{2}{Z^Y_{B_2Y_2}}}$ | $\frac{2q_B}{Z^B_{B_2Y_2}} = \frac{2q_Y}{Z^Y_{B_2Y_2}} = Q_{B_2Y_2}$ |
| $[ABX_2]_{quad}$ | $Z^A_{ABX_2}$ | $Z^B_{ABX_2}$ | $Z^X_{ABX_2}$ | $Z^X_{ABX_2}$ | $A_{\frac{1}{Z^A_{ABX_2}}} B_{\frac{1}{Z^B_{ABX_2}}} X_{\frac{2}{Z^X_{ABX_2}}}$ | $\frac{q_A}{Z^A_{ABX_2}} + \frac{q_B}{Z^B_{ABX_2}} = \frac{2q_X}{Z^X_{ABX_2}} = Q_{ABX_2}$ |
| $[ABY_2]_{quad}$ | $Z^A_{ABY_2}$ | $Z^B_{ABY_2}$ | $Z^Y_{ABY_2}$ | $Z^Y_{ABY_2}$ | $A_{\frac{1}{Z^A_{ABY_2}}} B_{\frac{1}{Z^B_{ABY_2}}} Y_{\frac{2}{Z^Y_{ABY_2}}}$ | $\frac{q_A}{Z^A_{ABY_2}} + \frac{q_B}{Z^B_{ABY_2}} = \frac{2q_Y}{Z^Y_{ABY_2}} = Q_{ABY_2}$ |
| $[A_2XY]_{quad}$ | $Z^A_{A_2XY}$ | $Z^A_{A_2XY}$ | $Z^X_{A_2XY}$ | $Z^Y_{A_2XY}$ | $A_{\frac{2}{Z^A_{A_2XY}}} X_{\frac{1}{Z^X_{A_2XY}}} Y_{\frac{1}{Z^Y_{A_2XY}}}$ | $\frac{2q_A}{Z^A_{A_2XY}} = \frac{q_X}{Z^X_{A_2XY}} + \frac{q_Y}{Z^Y_{A_2XY}} = Q_{A_2XY}$ |
| $[B_2XY]_{quad}$ | $Z^B_{B_2XY}$ | $Z^B_{B_2XY}$ | $Z^X_{B_2XY}$ | $Z^Y_{B_2XY}$ | $B_{\frac{2}{Z^B_{B_2XY}}} X_{\frac{1}{Z^X_{B_2XY}}} Y_{\frac{1}{Z^Y_{B_2XY}}}$ | $\frac{2q_B}{Z^B_{B_2XY}} = \frac{q_X}{Z^X_{B_2XY}} + \frac{q_Y}{Z^Y_{B_2XY}} = Q_{B_2XY}$ |
| $[ABXY]_{quad}$ | $Z^A_{ABXY}$ | $Z^B_{ABXY}$ | $Z^X_{ABXY}$ | $Z^Y_{ABXY}$ | $A_{\frac{1}{Z^A_{ABXY}}} B_{\frac{1}{Z^B_{ABXY}}} X_{\frac{1}{Z^X_{ABXY}}} Y_{\frac{1}{Z^Y_{ABXY}}}$ | $\frac{q_A}{Z^A_{ABXY}} + \frac{q_B}{Z^B_{ABXY}} = \frac{q_X}{Z^X_{ABXY}} + \frac{q_Y}{Z^Y_{ABXY}} = Q_{ABXY}$ |

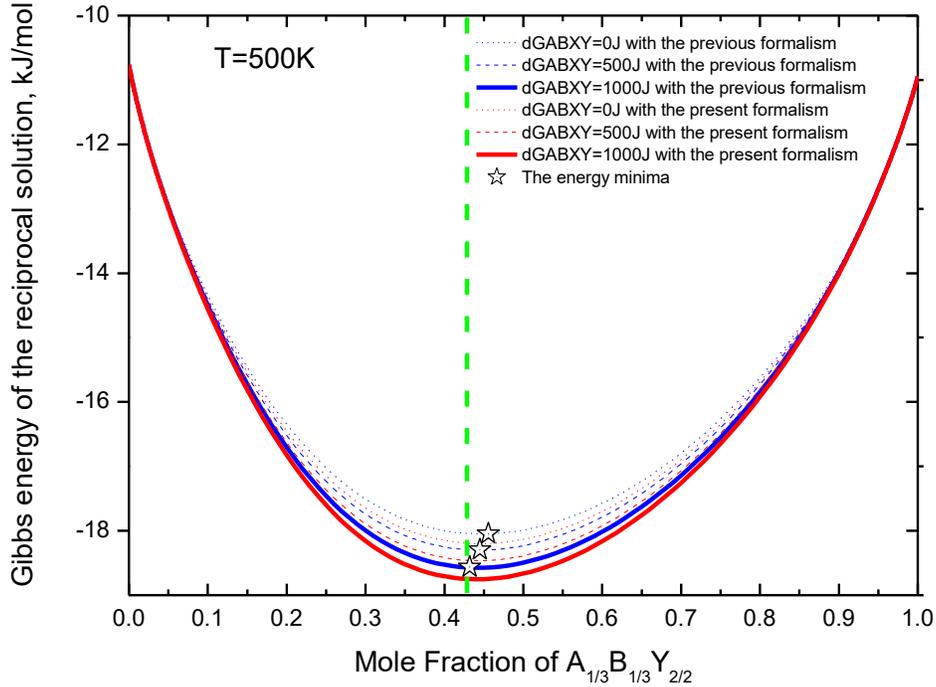

Fig. 3 Calculated energy of the reciprocal ABXY solution along the $A_{1/2}B_{1/4}X_{2/2}$-$A_{1/3}B_{1/3}Y_{2/2}$ line

In the next section, a suitable approach is proposed to develop the new energy formalism for reciprocal quadruplets. This method circumvents the mass-balance issue arising from the singular matrix. The derived formalism rigorously matches the mass conservation between the reciprocal quadruplet and all the constituent binary quadruplets without requiring adherence to equations (17-20) and only respecting the charge-neutrality condition for each quadruplet.

## 3. Model improvements

As for the aforementioned discussions, the previous formalism [11] can completely balance the quasichemical reaction equation between the ternary reciprocal quadruplet and all its surrounding binary quadruplets only if unified "charge-equivalent" fractions are imposed on each cation and each anion in the respective cationic and anionic sublattices over all the quadruplets, as reflected in equations (17-20). Due to the charge-neutrality condition, the matrix rank of equation (15) is less than the number of unknown variables, which results in an infinite number of solutions of the balancing coefficients. Essentially, from the four binary quadruplets, any three of them could be singled out to match the chemistry of the reciprocal quadruplet. In this way, the four binary quadruplets could be permutated into four groups ($mn\bar{o}$, $n\bar{o}p$, $m\bar{o}p$ and $mnp$). Each group contains three binary quadruplets. As exhibited in Fig. 1, $m$, $n$, $\bar{o}$ and $p$ refer to $[ABX_2]_{quad}$, $[B_2XY]_{quad}$, $[ABY_2]_{quad}$ and $[A_2XY]_{quad}$, respectively.

For the $mn\bar{o}$ group, the mass-balance equations are known in equations (22-25),

$$\frac{m_{mn\bar{o}}}{Z^A_{ABX_2}} + \frac{\bar{o}_{mn\bar{o}}}{Z^A_{ABY_2}} = \frac{1}{Z^A_{ABXY}} \tag{22}$$

$$\frac{2m_{mn\bar{o}}}{Z^X_{ABX_2}} + \frac{n_{mn\bar{o}}}{Z^X_{B_2XY}} = \frac{1}{Z^X_{ABXY}} \tag{23}$$

$$\frac{2\bar{o}_{mn\bar{o}}}{Z^Y_{ABY_2}} + \frac{n_{mn\bar{o}}}{Z^Y_{B_2XY}} = \frac{1}{Z^Y_{ABXY}} \tag{24}$$

$$\frac{m_{mn\bar{o}}}{Z^B_{ABX_2}} + \frac{\bar{o}_{mn\bar{o}}}{Z^B_{ABY_2}} + \frac{2n_{mn\bar{o}}}{Z^B_{B_2XY}} = \frac{1}{Z^B_{ABXY}} \tag{25}$$

where equation (25) is assigned as the dependent equation due to the charge-neutrality condition. The unknown variables $m_{mn\bar{o}}$, $n_{mn\bar{o}}$ and $\bar{o}_{mn\bar{o}}$ are easily solved by the following equivalent matrix,

$$\begin{bmatrix} \frac{1}{Z^A_{ABX_2}} & 0 & \frac{1}{Z^A_{ABY_2}} \\ \frac{2}{Z^X_{ABX_2}} & \frac{1}{Z^X_{B_2XY}} & 0 \\ 0 & \frac{1}{Z^Y_{B_2XY}} & \frac{2}{Z^Y_{ABY_2}} \end{bmatrix} \begin{bmatrix} m_{mn\bar{o}} \\ n_{mn\bar{o}} \\ \bar{o}_{mn\bar{o}} \end{bmatrix} = \begin{bmatrix} \frac{1}{Z^A_{ABXY}} \\ \frac{1}{Z^X_{ABXY}} \\ \frac{1}{Z^Y_{ABXY}} \end{bmatrix} \quad (26)$$

where $m_{mn\bar{o}}$, $n_{mn\bar{o}}$ and $\bar{o}_{mn\bar{o}}$ are associated with [ABX$_2$]$_{quad}$, [B$_2$XY]$_{quad}$ and [ABY$_2$]$_{quad}$ in the $mn\bar{o}$ group, respectively. In the same way, equations (27-31) are used to solve the corresponding coefficients for [B$_2$XY]$_{quad}$, [ABY$_2$]$_{quad}$ and [A$_2$XY]$_{quad}$ in the $n\bar{o}p$ group,

$$\frac{\bar{o}_{n\bar{o}p}}{Z^A_{ABY_2}} + \frac{2p_{n\bar{o}p}}{Z^A_{A_2XY}} = \frac{1}{Z^A_{ABXY}} \quad (27)$$

$$\frac{2n_{n\bar{o}p}}{Z^B_{B_2XY}} + \frac{\bar{o}_{n\bar{o}p}}{Z^B_{ABY_2}} = \frac{1}{Z^B_{ABXY}} \quad (28)$$

$$\frac{n_{n\bar{o}p}}{Z^X_{B_2XY}} + \frac{p_{n\bar{o}p}}{Z^X_{A_2XY}} = \frac{1}{Z^X_{ABXY}} \quad (29)$$

$$\frac{n_{n\bar{o}p}}{Z^Y_{B_2XY}} + \frac{2\bar{o}_{n\bar{o}p}}{Z^Y_{ABY_2}} + \frac{p_{n\bar{o}p}}{Z^Y_{A_2XY}} = \frac{1}{Z^Y_{ABXY}} \quad (30)$$

$$\begin{bmatrix} 0 & \frac{1}{Z^A_{ABY_2}} & \frac{2}{Z^A_{A_2XY}} \\ \frac{2}{Z^B_{B_2XY}} & \frac{1}{Z^B_{ABY_2}} & 0 \\ \frac{1}{Z^X_{B_2XY}} & 0 & \frac{1}{Z^X_{A_2XY}} \end{bmatrix} \begin{bmatrix} n_{n\bar{o}p} \\ o_{n\bar{o}p} \\ p_{n\bar{o}p} \end{bmatrix} = \begin{bmatrix} \frac{1}{Z^A_{ABXY}} \\ \frac{1}{Z^B_{ABXY}} \\ \frac{1}{Z^X_{ABXY}} \end{bmatrix} \quad (31)$$

where equation (30) is assigned to be dependent. Equations (32-36) are employed to solve those coefficients for [ABX$_2$]$_{quad}$, [ABY$_2$]$_{quad}$ and [A$_2$XY]$_{quad}$ in the $m\bar{o}p$ group,

$$\frac{m_{m\bar{o}p}}{Z^B_{ABX_2}} + \frac{\bar{o}_{m\bar{o}p}}{Z^B_{ABY_2}} = \frac{1}{Z^B_{ABXY}} \quad (32)$$

$$\frac{2m_{m\bar{o}p}}{Z^X_{ABX_2}} + \frac{p_{m\bar{o}p}}{Z^X_{A_2XY}} = \frac{1}{Z^X_{ABXY}} \quad (33)$$

$$\frac{2\bar{o}_{m\bar{o}p}}{Z^Y_{ABY_2}} + \frac{p_{m\bar{o}p}}{Z^Y_{A_2XY}} = \frac{1}{Z^Y_{ABXY}} \quad (34)$$

$$\frac{\bar{o}_{m\bar{o}p}}{Z^A_{ABY_2}} + \frac{2p_{m\bar{o}p}}{Z^A_{A_2XY}} + \frac{m_{m\bar{o}p}}{Z^A_{ABX_2}} = \frac{1}{Z^A_{ABXY}} \quad (35)$$

$$\begin{bmatrix} \frac{1}{Z^B_{ABX_2}} & \frac{1}{Z^B_{ABY_2}} & 0 \\ \frac{2}{Z^X_{ABX_2}} & 0 & \frac{1}{Z^X_{A_2XY}} \\ 0 & \frac{2}{Z^Y_{ABY_2}} & \frac{1}{Z^Y_{A_2XY}} \end{bmatrix} \begin{bmatrix} m_{m\bar{o}p} \\ \bar{o}_{m\bar{o}p} \\ p_{m\bar{o}p} \end{bmatrix} = \begin{bmatrix} \frac{1}{Z^B_{ABXY}} \\ \frac{1}{Z^X_{ABXY}} \\ \frac{1}{Z^Y_{ABXY}} \end{bmatrix} \quad (36)$$

where equation (35) is assigned to be dependent. Equations (37-41) are used to determine the coefficients for [ABX$_2$]$_{quad}$, [A$_2$XY]$_{quad}$ and [B$_2$XY]$_{quad}$ in the *mnp* group,

$$\frac{m_{mnp}}{Z^A_{ABX_2}} + \frac{2p_{mnp}}{Z^A_{A_2XY}} = \frac{1}{Z^A_{ABXY}} \quad (37)$$

$$\frac{m_{mnp}}{Z^B_{ABX_2}} + \frac{2n_{mnp}}{Z^B_{B_2XY}} = \frac{1}{Z^B_{ABXY}} \quad (38)$$

$$\frac{n_{mnp}}{Z^Y_{B_2XY}} + \frac{p_{mnp}}{Z^Y_{A_2XY}} = \frac{1}{Z^Y_{ABXY}} \quad (39)$$

$$\frac{p_{mnp}}{Z^X_{A_2XY}} + \frac{2m_{mnp}}{Z^X_{ABX_2}} + \frac{n_{mnp}}{Z^X_{B_2XY}} = \frac{1}{Z^X_{ABXY}} \quad (40)$$

$$\begin{bmatrix} \frac{1}{Z^A_{ABX_2}} & 0 & \frac{2}{Z^A_{A_2XY}} \\ \frac{1}{Z^B_{ABX_2}} & \frac{2}{Z^B_{B_2XY}} & 0 \\ 0 & \frac{1}{Z^Y_{B_2XY}} & \frac{1}{Z^Y_{A_2XY}} \end{bmatrix} \begin{bmatrix} m_{mnp} \\ n_{mnp} \\ p_{mnp} \end{bmatrix} = \begin{bmatrix} \frac{1}{Z^A_{ABXY}} \\ \frac{1}{Z^B_{ABXY}} \\ \frac{1}{Z^Y_{ABXY}} \end{bmatrix} \quad (41)$$

where equation (40) is assigned to be dependent. So far, there are three coefficients determined for each of the four binary quadruplets from equations (22-41). The coefficients $m_{mn\bar{o}}$, $m_{m\bar{o}p}$ and $m_{mnp}$ are associated with [ABX$_2$]$_{quad}$; $n_{mn\bar{o}}$, $n_{n\bar{o}p}$ and $n_{mnp}$ with [B$_2$XY]$_{quad}$; $\bar{o}_{mn\bar{o}}$, $\bar{o}_{n\bar{o}p}$ and $\bar{o}_{m\bar{o}p}$ with [ABY$_2$]$_{quad}$; and $p_{n\bar{o}p}$, $p_{m\bar{o}p}$ and $p_{mnp}$ with [A$_2$XY]$_{quad}$. From observation of equations (22-41), it is seen that they can be classified into four groups: equations (22, 27, 35, 37), equations (25, 28, 32, 38), equations (23, 29, 33, 39), and equations (24, 30, 34, 39). All the equations in each group can be merged to generate equations (42-45),

$$\frac{(m_{mn\bar{o}}+m_{m\bar{o}p}+m_{mnp})}{Z^A_{ABX_2}} + \frac{(\bar{o}_{mn\bar{o}}+\bar{o}_{n\bar{o}p}+\bar{o}_{m\bar{o}p})}{Z^A_{ABY_2}} + \frac{2(p_{n\bar{o}p}+p_{m\bar{o}p}+p_{mnp})}{Z^A_{A_2XY}} = \frac{4}{Z^A_{ABXY}} \quad (42)$$

$$\frac{(m_{mn\bar{o}}+m_{m\bar{o}p}+m_{mnp})}{Z^B_{ABX_2}} + \frac{2(n_{mn\bar{o}}+n_{n\bar{o}p}+n_{mnp})}{Z^B_{B_2XY}} + \frac{(\bar{o}_{mn\bar{o}}+\bar{o}_{n\bar{o}p}+\bar{o}_{m\bar{o}p})}{Z^B_{ABY_2}} = \frac{4}{Z^B_{ABXY}} \quad (43)$$

$$\frac{2(m_{mn\bar{o}}+m_{m\bar{o}p}+m_{mnp})}{Z^X_{ABX_2}} + \frac{(n_{mn\bar{o}}+n_{n\bar{o}p}+n_{mnp})}{Z^X_{B_2XY}} + \frac{(p_{n\bar{o}p}+p_{m\bar{o}p}+p_{mnp})}{Z^X_{A_2XY}} = \frac{4}{Z^X_{ABXY}} \quad (44)$$

$$\frac{(n_{mn\bar{o}}+n_{n\bar{o}p}+n_{mnp})}{Z^Y_{B_2XY}} + \frac{2(\bar{o}_{mn\bar{o}}+\bar{o}_{n\bar{o}p}+\bar{o}_{m\bar{o}p})}{Z^Y_{ABY_2}} + \frac{(p_{n\bar{o}p}+p_{m\bar{o}p}+p_{mnp})}{Z^Y_{A_2XY}} = \frac{4}{Z^Y_{ABXY}} \quad (45)$$

where each equation is inherently correct and stands strictly for the mass balance between the reciprocal quadruplet and all the constituent binary quadruplets. By comparing equations (42-45) to equations (11-14), the following expressions are defined,

$$m = (m_{mn\bar{o}} + m_{m\bar{o}p} + m_{mnp}) \qquad [ABX_2]_{quad} \quad (46)$$

$$n = (n_{mn\bar{o}} + n_{n\bar{o}p} + n_{mnp}) \qquad [B_2XY]_{quad} \quad (47)$$

$$\bar{o} = (\bar{o}_{mn\bar{o}} + \bar{o}_{n\bar{o}p} + \bar{o}_{m\bar{o}p}) \qquad [ABY_2]_{quad} \quad (48)$$

$$p = (p_{n\bar{o}p} + p_{m\bar{o}p} + p_{mnp}) \qquad [A_2XY]_{quad} \quad (49)$$

where the balance coefficients $m$, $n$, $\bar{o}$ and $p$ are finally found for the four binary $[ABX_2]_{quad}$, $[ABY_2]_{quad}$, $[B_2XY]_{quad}$ and $[A_2XY]_{quad}$, respectively.

Let us compare the difference and similarity between the two formalisms. When the reciprocal quadruplet $A_{\frac{1}{4}}B_{\frac{1}{6}}X_{\frac{1}{3}}Y_{\frac{1}{4}}$ and all the binary quadruplets $A_{\frac{1}{2}}B_{\frac{1}{6}}X_{\frac{2}{2.4}}$, $A_{\frac{1}{6}}B_{\frac{1}{3}}Y_{\frac{2}{2.4}}$, $A_{\frac{2}{4}}X_{\frac{1}{3}}Y_{\frac{1}{6}}$ and $B_{\frac{2}{6}}X_{\frac{1}{2}}Y_{\frac{1}{6}}$ are defined, then the $m$, $n$, $\bar{o}$, $p$ calculated by the formalism of Pelton et al. [11] are 0.8, 1.0, 0.6 and 1.0 compared to 0.4667, 1.0303, 0.7364 and 1.2879 from the new formalism. The new formalism rigorously complies with the mass-conservation constraints while the old one does not. Furthermore, the previous formalism imposes the condition that all balance coefficients must be positive, which, in reality, fails to describe a potential miscibility gap between the reciprocal quadruplet and some of the binary quadruplets. Lambotte et al. [30] modeled $Na_2O$-$SiO_2$-$NaF$-$SiF_4$ reciprocal liquids where the quasichemical reaction are defined as,

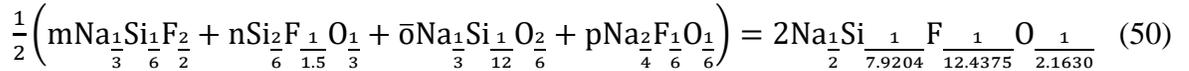
$$\frac{1}{2}\left(mNa_{\frac{1}{3}}Si_{\frac{1}{6}}F_{\frac{2}{2}} + nSi_{\frac{2}{6}}F_{\frac{1}{1.5}}O_{\frac{1}{3}} + \bar{o}Na_{\frac{1}{3}}Si_{\frac{1}{12}}O_{\frac{2}{6}} + pNa_{\frac{2}{4}}F_{\frac{1}{6}}O_{\frac{1}{6}}\right) = 2Na_{\frac{1}{2}}Si_{\frac{1}{7.9204}}F_{\frac{1}{12.4375}}O_{\frac{1}{2.1630}} \quad (50)$$

and $m$, $n$, $\bar{o}$ and $p$ are determined to be 0.1608, 0.7575, 2.7739 and 2 by the previous formalism, leading to the ions of Na, Si, F and O on the left side of equation (50) as 1/2.0020, 1/7.8359, 1/4.0034 and 1/2.6482 compared to those on the right side as 1/2, 1/7.9204, 1/12.4375 and 1/2.1630, respectively. Using the new formalism, $m$, $n$, $\bar{o}$ and $p$ are calculated to be -0.4425, 0.7380, 3.9935 and 1.6327, causing mass balance constraints to be respected between the two sides of equation (50). Here, the negative coefficient $m$ indicates that the binary quadruplet $Na_{\frac{1}{3}}Si_{\frac{1}{6}}F_{\frac{2}{2}}$ could be displaced from the left side to the right side of the quasichemical reaction. When $\Delta G_{NaSiFO}$ becomes negative, the reaction direction will be shifted toward the right side of equation (50); [NaSiF$_2$]$_{quad}$ and [NaSiFO]$_{quad}$ tend to dominate the reciprocal solution and a miscibility gap is thus formed along the [NaSiF$_2$]$_{quad}$-[NaSiFO]$_{quad}$ line. Most importantly, the calculated energies using the new formalism are distinct from what the previous formalism produced. As shown from the red curves in Fig. 3, the new formalism predicts the energy minimum lying as expected at the position of the reciprocal quadruplet. If all the coordination numbers are defined according to equations (17-20), then the new formalism will be equivalent to the previous one of Pelton et al. [11]. Taking the coefficient $m$ to be solved for example, the matrix equations (26, 31, 36, 41) could be equivalently transformed to the following ones if equations (17-20) must be respected:

$$\begin{bmatrix} Q_{ABX_2}Y'_A & 0 & Q_{ABY_2}Y'_A \\ Q_{ABX_2} & Q_{B_2XY}Y'_X & 0 \\ 0 & Q_{B_2XY}Y'_Y & Q_{ABY_2} \end{bmatrix} \begin{bmatrix} m_{mn\bar{o}} \\ n_{mn\bar{o}} \\ \bar{o}_{mn\bar{o}} \end{bmatrix} = \begin{bmatrix} Q_{ABXY}Y'_A \\ Q_{ABXY}Y'_X \\ Q_{ABXY}Y'_Y \end{bmatrix} \quad (51)$$

$$\begin{bmatrix} Q_{ABX_2}Y'_B & Q_{ABY_2}Y'_B & 0 \\ Q_{ABX_2} & 0 & Q_{A_2XY}Y'_X \\ 0 & Q_{ABY_2} & Q_{A_2XY}Y'_Y \end{bmatrix} \begin{bmatrix} m_{m\bar{o}p} \\ \bar{o}_{m\bar{o}p} \\ p_{m\bar{o}p} \end{bmatrix} = \begin{bmatrix} Q_{ABXY}Y'_B \\ Q_{ABXY}Y'_X \\ Q_{ABXY}Y'_Y \end{bmatrix} \quad (52)$$

$$\begin{bmatrix} Q_{ABX_2}Y'_A & 0 & Q_{A_2XY} \\ Q_{ABX_2}Y'_B & Q_{B_2XY} & 0 \\ 0 & Q_{B_2XY}Y'_Y & Q_{A_2XY}Y'_Y \end{bmatrix} \begin{bmatrix} m_{mnp} \\ n_{mnp} \\ p_{mnp} \end{bmatrix} = \begin{bmatrix} Q_{ABXY}Y'_A \\ Q_{ABXY}Y'_B \\ Q_{ABXY}Y'_Y \end{bmatrix} \quad (53)$$

$$\begin{bmatrix} 0 & Q_{ABY_2}Y'_A & Q_{A_2XY} \\ Q_{B_2XY} & Q_{ABY_2}Y'_B & 0 \\ Q_{B_2XY}Y'_X & 0 & Q_{A_2XY}Y'_X \end{bmatrix} \begin{bmatrix} n_{n\bar{o}p} \\ \bar{o}_{n\bar{o}p} \\ p_{n\bar{o}p} \end{bmatrix} = \begin{bmatrix} Q_{ABXY}Y'_A \\ Q_{ABXY}Y'_B \\ Q_{ABXY}Y'_X \end{bmatrix} \quad (54)$$

where $Q_{ijkl}$ is the total charge of the cationic or anionic sublattice for the [ijkl]$_{quad}$. For the [ABX$_2$]$_{quad}$, $Q_{ABX_2} = \frac{2q_X}{Z^X_{ABX_2}} = \frac{q_A}{Z^A_{ABX_2}} + \frac{q_B}{Z^B_{ABX_2}}$ is defined, and all the other quadruplets have similar forms. From equations (51-53), $m_{mn\bar{o}} = \frac{Q_{ABXY}Y'_X}{Q_{ABX_2}}$, $m_{m\bar{o}p} = \frac{Q_{ABXY}Y'_X}{Q_{ABX_2}}$, and $m_{mnp} = 0$ are determined, and $\frac{Q_{ABXY}Y'_X}{Q_{ABX_2}}$ is equivalent to $\frac{Z^X_{ABX_2}}{2Z^X_{ABXY}}$ as shown below,

$$\frac{Q_{ABXY}Y'_X}{Q_{ABX_2}} = \frac{\left(\frac{q_X}{Z^X_{ABXY}} + \frac{q_Y}{Z^X_{ABXY}}\right) \cdot \frac{\frac{q_X}{Z^X_{ABXY}}}{\frac{q_X}{Z^X_{ABXY}} + \frac{q_Y}{Z^X_{ABXY}}}}{2\frac{q_X}{Z^X_{ABX_2}}} = \frac{Z^X_{ABX_2}}{2Z^X_{ABXY}} \tag{55}$$

leading to

$$m = m_{mn\bar{o}} + m_{m\bar{o}p} + m_{mnp} = \frac{Z^X_{ABX_2}}{Z^X_{ABXY}} \tag{56}$$

which is exactly the same as that proposed by Pelton et al. [11]. The similar situation could be found for all the other coefficients $n$, $\bar{o}$, $p$ as well. The new formalism is generic without any constraints other than the charge-neutrality condition for all the quadruplets.

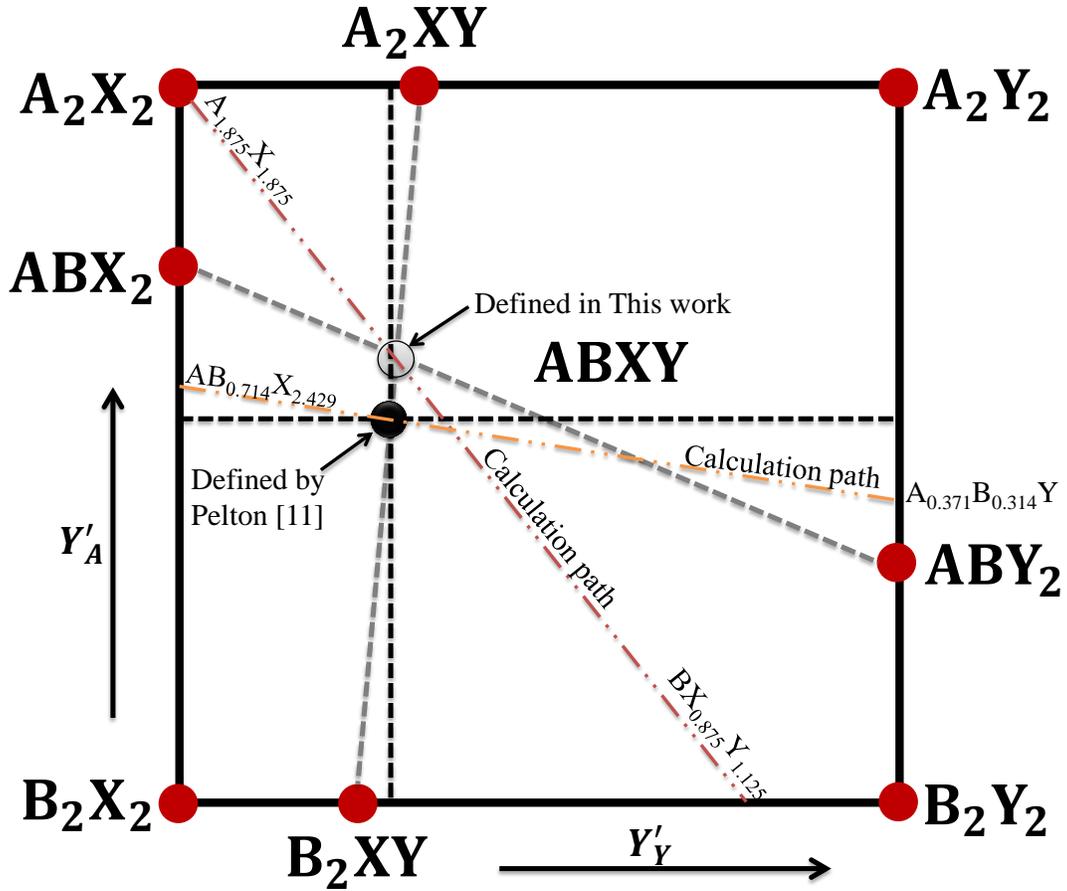

Fig. 4 Default positions of the quadruplet ABXY defined from Pelton [11] and the current work ($Y'_A$ and $Y'_Y$ are defined by eqs. 17 and 20, respectively)

On most occasions, there is no clear experimental evidence to confirm the existence of SRO in reciprocal solutions. Hence, the default coordination numbers $Z^i_{ABXY}$ and the resultant chemistry for the reciprocal quadruplet were previously defined by Pelton et al. [11] by the following equations:

$$\frac{1}{Z^A_{ABXY}} = (\frac{Z^X_{ABX_2}}{q_X Z^A_{ABX_2}} + \frac{Z^Y_{ABY_2}}{q_Y Z^A_{ABY_2}})F \tag{57}$$

$$\frac{1}{Z^X_{ABXY}} = (\frac{Z^A_{A_2XY}}{q_A Z^X_{A_2XY}} + \frac{Z^B_{B_2XY}}{q_B Z^X_{B_2XY}})F \tag{58}$$

and similarly for $Z^B_{ABXY}$ and $Z^Y_{ABXY}$, where

$$F = \frac{1}{8}(\frac{q_X}{Z^X_{ABX_2}} + \frac{q_Y}{Z^Y_{ABY_2}} + \frac{q_A}{Z^A_{A_2XY}} + \frac{q_B}{Z^B_{B_2XY}}) \tag{59}$$

This definition was based on the assumption that the position of the [ABXY]$_{quad}$ lay at the average of the values of $Y'_A$ of the quadruplets [ABX$_2$]$_{quad}$ and [ABY$_2$]$_{quad}$ and at the average of the values of $Y'_Y$ of the quadruplets [A$_2$XY]$_{quad}$ and [B$_2$XY]$_{quad}$, as illustrated by the black point in Fig. 4. The present work proposes a different approach where the position is defined as the intersection point (the grey point in Fig. 4) of the ABX$_2$-ABY$_2$ line and A$_2$XY-B$_2$XY line. The coordinates of this point can be determined by solving the equations of the two straight lines and thus by the following matrix,

$$\begin{bmatrix} Y'_{A(m)} - Y'_{A(\bar{o})} & 1 \\ 1 & Y'_{Y(n)} - Y'_{Y(p)} \end{bmatrix} \begin{bmatrix} Y'_Y \\ Y'_A \end{bmatrix} = \begin{bmatrix} Y'_{A(m)} \\ Y'_{Y(n)} \end{bmatrix} \tag{60}$$

where the coordinates' vector is solved as below,

$$Y'_Y = \frac{Y'_{Y(n)} + Y'_{Y(p)} Y'_{A(m)} - Y'_{A(m)} Y'_{Y(n)}}{Y'_{A(m)} Y'_{Y(p)} - Y'_{A(m)} Y'_{Y(n)} + Y'_{A(\bar{o})} Y'_{Y(n)} - Y'_{A(\bar{o})} Y'_{Y(p)} + 1} \qquad Y'_X = 1 - Y'_Y \tag{61}$$

$$Y'_A = \frac{Y'_{A(m)} + Y'_{A(\bar{o})} Y'_{Y(n)} - Y'_{A(m)} Y'_{Y(n)}}{Y'_{A(m)} Y'_{Y(p)} - Y'_{A(m)} Y'_{Y(n)} + Y'_{A(\bar{o})} Y'_{Y(n)} - Y'_{A(\bar{o})} Y'_{Y(p)} + 1} \qquad Y'_B = 1 - Y'_A \tag{62}$$

with $m$, $n$, $\bar{o}$ and $p$ representing the corresponding binary quadruplets and the "charge-equivalent" fractions calculated by equations (17-20). The default coordination numbers defined in this work are expressed as,

$$\frac{1}{Z^i_{ABXY}} = \frac{Q'_{ABXY} Y'_i}{q_i} \tag{63}$$

where

$$Q'_{ABXY} = \frac{1}{2}(\frac{q_X}{Z^X_{ABX_2}} + \frac{q_Y}{Z^Y_{ABY_2}} + \frac{q_A}{Z^A_{A_2XY}} + \frac{q_B}{Z^B_{B_2XY}}) \tag{64}$$

It is obvious that the two definitions regarding the default coordination numbers are the same when equations (17-20) are observed. Moreover, by using the present definition, equation (16) from Pelton et al. [11] will become completely correct, which means the mass balance is strictly respected between the "reactants" and the "products". The binary quadruplets $A_{\frac{1}{2}}B_{\frac{1}{6}}X_{\frac{2}{2.4}}$, $A_{\frac{1}{6}}B_{\frac{1}{3}}Y_{\frac{2}{2.4}}$, $A_{\frac{2}{4}}X_{\frac{1}{3}}Y_{\frac{1}{6}}$ and $B_{\frac{2}{6}}X_{\frac{1}{2}}Y_{\frac{1}{6}}$ are still used here to compare the calculated results between the previous and present definitions. The former generates the reciprocal quadruplet $A_{\frac{1}{3.529}}B_{\frac{1}{4.706}}X_{\frac{1}{1.993}}Y_{\frac{1}{4.840}}$ while the latter creates $A_{\frac{1}{2.918}}B_{\frac{1}{5.471}}X_{\frac{1}{1.989}}Y_{\frac{1}{4.863}}$. The $AB_{0.714}X_{2.429}$-$A_{0.371}B_{0.314}Y$ and $A_{1.875}X_{1.875}$-$BX_{0.875}Y_{1.125}$ (seeing the yellow and red dash-dot-dot lines in Fig.4) lines are designed in the reciprocal composition space so as to have the former and latter quadruplets correctly located at the intersection of the two respective lines. Fig.5 displays the calculated energies of the reciprocal solution along the two calculation paths. As can be seen, the curve minima are located near the connect composition of the latter quadruplet while they drift a bit away from that of the former quadruplet. The accordance and deviation certainly result from the definitions of the default coordination numbers and the default Gibbs energies of the reciprocal quadruplet.

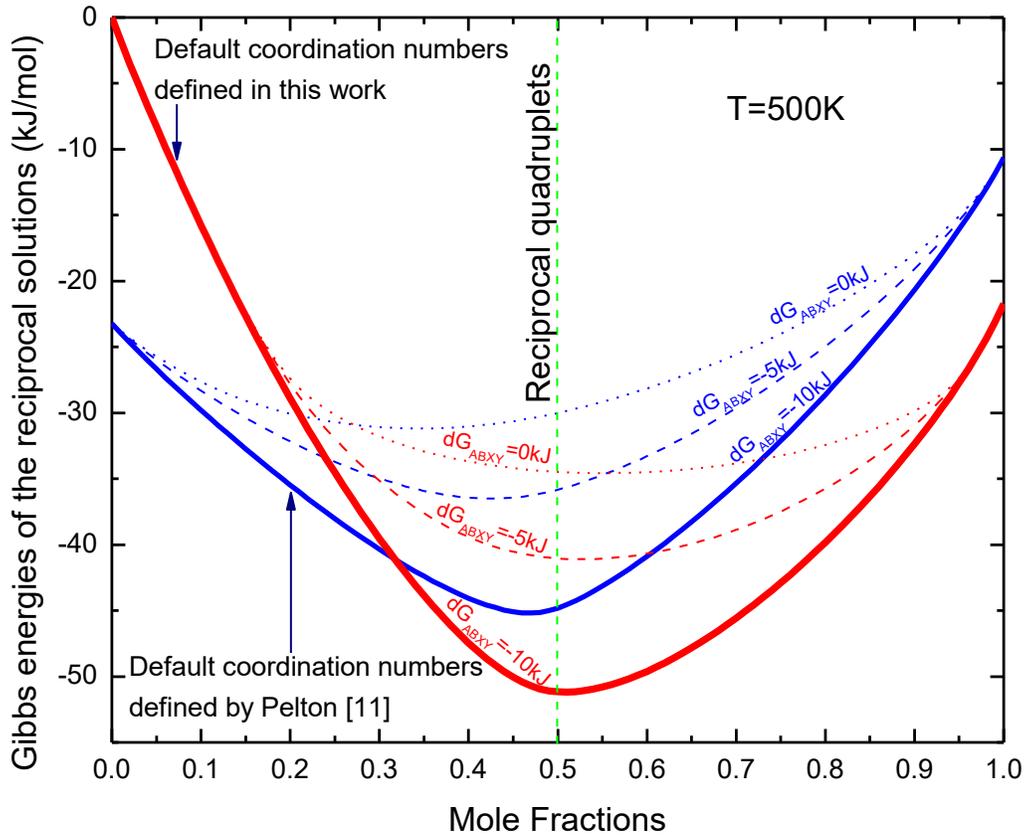

Fig. 5 Calculated energy of the reciprocal ABXY solution using the default coordination numbers defined by Pelton [11] along the $AB_{0.714}X_{2.429}$-$A_{0.371}B_{0.314}Y$ line and in this work along the $A_{1.875}X_{1.875}$-$BX_{0.875}Y_{1.125}$ line

It is noteworthy that, no matter what the coordination numbers of the reciprocal quadruplet are, the balancing coefficients can be uniquely determined by equations (46-49) with rigorous mass conservation. The Appendix lists the analytical forms for all the balance coefficients in order to facilitate their code implementation. As discussed above, any three binary quadruplets could make up one reciprocal quadruplet. However, the default Gibbs energy of the reciprocal quadruplet cannot be reasonably defined using only three constituent quadruplets, since the quasichemical reaction in a reciprocal solution is supposed to occur between the reciprocal quadruplet and all its surrounding quadruplets. Furthermore, assembling the reciprocal quadruplet using three binary quadruplets might cause the energy landscape of the reciprocal solution to lose symmetry and balance. Based on equations (10, 46-49), the formalism for the standard Gibbs energy of [ABXY]$_{quad}$ is derived as,

$$G^0_{ABXY} = \frac{1}{4}(mG^0_{ABX_2} + nG^0_{B_2XY} + \bar{o}G^0_{ABY_2} + pG^0_{A_2XY}) + \Delta G^0_{ABXY} \qquad (65)$$

where $G^0_{ABX_2}$ was previously defined as equation (7), and all the other binary quadruplets are described with similar energy formalisms. Herein, all the nine quadruplets in the reciprocal ABXY solution are well defined, and equation (1) is rewritten as follows,

$$G = n_{A_2X_2}G^0_{A_2X_2} + n_{B_2X_2}G^0_{B_2X_2} + n_{A_2Y_2}G^0_{A_2Y_2} + n_{B_2Y_2}G^0_{B_2Y_2} + n_{ABX_2}G^0_{ABX_2} + n_{ABY_2}G^0_{ABY_2} +$$
$$n_{A_2XY}G^0_{A_2XY} + n_{B_2XY}G^0_{B_2XY} + n_{ABXY}G^0_{ABXY} + \left(n_{ABX_2} + \frac{m}{4}n_{ABXY}\right)\left(\Delta G_{ABX_2} - \Delta G^0_{ABX_2}\right) +$$
$$\left(n_{ABY_2} + \frac{\bar{o}}{4}n_{ABXY}\right)\left(\Delta G_{ABY_2} - \Delta G^0_{ABY_2}\right) + \left(n_{A_2XY} + \frac{p}{4}n_{ABXY}\right)\left(\Delta G_{A_2XY} - \Delta G^0_{A_2XY}\right) +$$
$$\left(n_{B_2XY} + \frac{n}{4}n_{ABXY}\right)\left(\Delta G_{B_2XY} - \Delta G^0_{B_2XY}\right) + n_{ABXY}\left(\Delta G_{ABXY} - \Delta G^0_{ABXY}\right) - T\Delta S^{config} \qquad (66)$$

where all $\Delta G_{ijkl}$ are empirical parameters to be optimized by available experimental phase equilibria and thermochemical data, and the configurational entropy $\Delta S_{config}$ also takes on a form originally proposed by Pelton et al. [11], and later improved by Pelton et al. [31], for strong FNN short-range ordering. The improved expression of $\Delta S_{config}$ is presented as,

$$-\frac{\Delta S^{config}}{R} = (n_A \ln X_A + n_B \ln X_B + n_X \ln X_X + n_Y \ln X_Y) + \left[n^*_{AX} \ln\left(\frac{X^*_{AX}}{F_AF_X}\right) + n^*_{BX} \ln\left(\frac{X^*_{BX}}{F_BF_X}\right) + \right.$$
$$\left. n^*_{AY} \ln\left(\frac{X^*_{AY}}{F_AF_Y}\right) + n^*_{BY} \ln\left(\frac{X^*_{BY}}{F_BF_Y}\right)\right] + \left[n_{A_2X_2} \ln\left(\frac{X_{A_2X_2}}{X^3_{AX}}Y_AY_X\right) + n_{B_2X_2} \ln\left(\frac{X_{B_2X_2}}{X^3_{BX}}Y_BY_X\right) + \right.$$
$$n_{A_2Y_2} \ln\left(\frac{X_{A_2Y_2}}{X^3_{AY}}Y_AY_Y\right) + n_{B_2Y_2} \ln\left(\frac{X_{B_2Y_2}}{X^3_{BY}}Y_BY_Y\right) + n_{ABX_2} \ln\left(\frac{X_{ABX_2}}{2X^{\frac{3}{2}}_{AX}X^{\frac{3}{2}}_{BX}}Y^{\frac{1}{2}}_AY^{\frac{1}{2}}_BY_X\right) +$$

$$n_{ABY_2} \ln\left(\frac{X_{ABY_2}}{2X_{AY}^{\frac{3}{2}}X_{BY}^{\frac{3}{2}}} Y_A^{\frac{1}{2}}Y_B^{\frac{1}{2}}Y_Y\right) + n_{A_2XY} \ln\left(\frac{X_{A_2XY}}{2X_{AX}^{\frac{3}{2}}X_{AY}^{\frac{3}{2}}} Y_A Y_X^{\frac{1}{2}}Y_Y^{\frac{1}{2}}\right) + n_{B_2XY} \ln\left(\frac{X_{B_2XY}}{2X_{BX}^{\frac{3}{2}}X_{BY}^{\frac{3}{2}}} Y_B Y_X^{\frac{1}{2}}Y_Y^{\frac{1}{2}}\right) +$$

$$n_{ABXY} \ln\left(\frac{X_{ABXY}}{4X_{AX}^{\frac{3}{4}}X_{BX}^{\frac{3}{4}}X_{AY}^{\frac{3}{4}}X_{BY}^{\frac{3}{4}}} Y_A^{\frac{1}{2}}Y_B^{\frac{1}{2}}Y_X^{\frac{1}{2}}Y_Y^{\frac{1}{2}}\right) \tag{67}$$

where $n_A$ is the number of moles of ion A, $X_A$ is the mole fraction of A, $n_{AX}^*$ stands for the number of moles of AX (the FNN pair) regarding the individual $\zeta$, $X_{AX}$ and $X_{AX}^*$ are the pair fractions of AX regarding the respective individual and unified $\zeta$, $Y_A$ and $F_A$ are the "coordination-equivalent" fractions of ion A regarding the respective individual and unified $\zeta$, $n_{A_2X_2}$ is the number of moles of [A$_2$X$_2$]$_{quad}$, $X_{A_2X_2}$ is the quadruplet fraction of [A$_2$X$_2$]$_{quad}$ (similarly for the others). These quantities and their relations were explicitly defined by Pelton et al. [11, 31]. The Appendix also reproduces their definitions and relations, and demonstrates the derivation of the configurational entropy. The internal variables $n_{ijkl}$ are determined by minimizing the Gibbs energy of the system at a given temperature and composition. They represent the internal equilibrium configuration where various types of SRO are characterized. It should be noted here that the newly proposed formalism will never affect calculations of any common-ion system and some specific reciprocal systems with the coordination numbers defined according to equations (17-20) since it is mainly developed for any general expression of the Gibbs energies of reciprocal quadruplets.

## 4. Concluding remarks

The current paper developed a new and generic formalism for expressing the default Gibbs energy of a reciprocal quadruplet within the two-sublattice quasichemical model. The new formalism maintains strict balance of the quasichemical reaction equation between the reciprocal quadruplet and all the surrounding binary quadruplets, and thus overcomes the previous imperfect definition of the default Gibbs energy of the reciprocal quadruplet in some limiting circumstances. This new progress is important to the MQMQA in terms of both theoretical rigorousness and practical applications. With the new development, it is expected to have more accurate thermodynamic predictions of reciprocal solutions. In other words, without any ternary reciprocal interaction parameters, the new formalism should be better able to perform credible simulations of thermodynamic properties and phase equilibria for general reciprocal solutions. With the newly developed formalism and all other recent improvements, the MQMQA is one of the most reliable thermodynamic models for various types of solutions with or without short-range order.

# Appendix

The balancing coefficients describing the mass-balance relations between reciprocal quadruplets and all the constituent binary quadruplets are derived and their analytic forms are presented here. These equations have been double checked for correctness by Matlab, but it is very difficult to simplify these equations to more concise forms. The Readers are welcome to conduct such simplifications if they are interested.

$m = (Z_{ABX_2}^B (Z_{A_2XY}^X Z_{ABX_2}^X Z_{ABY_2}^Y Z_{ABXY}^B Z_{ABXY}^Y + 2 Z_{A_2XY}^Y Z_{ABY_2}^B Z_{ABX_2}^X Z_{ABXY}^X Z_{ABXY}^Y - Z_{A_2XY}^Y Z_{ABX_2}^X Z_{ABY_2}^Y Z_{ABXY}^B Z_{ABXY}^X))/(8 Z_{ABXY}^B Z_{ABXY}^X Z_{ABXY}^Y (Z_{A_2XY}^X Z_{ABX_2}^B Z_{ABY_2}^Y + Z_{A_2XY}^Y Z_{ABY_2}^B Z_{ABX_2}^X)) + (Z_{ABX_2}^A (Z_{A_2XY}^A Z_{ABX_2}^B Z_{ABXY}^B Z_{ABXY}^Y Z_{B_2XY}^Y - 2 Z_{A_2XY}^Y Z_{ABX_2}^B Z_{ABXY}^A Z_{ABXY}^B Z_{B_2XY}^Y + Z_{A_2XY}^Y Z_{ABX_2}^B Z_{ABXY}^A Z_{ABXY}^Y Z_{B_2XY}^B))/(4 Z_{ABXY}^A Z_{ABXY}^B Z_{ABXY}^Y (Z_{A_2XY}^A Z_{ABX_2}^B Z_{B_2XY}^Y + Z_{A_2XY}^Y Z_{ABX_2}^A Z_{B_2XY}^B)) + (Z_{ABX_2}^A (2 Z_{ABY_2}^A Z_{ABX_2}^X Z_{ABXY}^X Z_{ABXY}^Y Z_{B_2XY}^Y - Z_{ABX_2}^X Z_{ABY_2}^Y Z_{ABXY}^A Z_{ABXY}^X Z_{B_2XY}^Y + Z_{ABX_2}^X Z_{ABY_2}^Y Z_{ABXY}^A Z_{ABXY}^Y Z_{B_2XY}^X))/(8 Z_{ABXY}^A Z_{ABXY}^X Z_{ABXY}^Y (Z_{ABX_2}^A Z_{ABY_2}^Y Z_{B_2XY}^Y + Z_{ABY_2}^A Z_{ABX_2}^X Z_{B_2XY}^Y))$  (A1)

$n = (Z_{B_2XY}^X (Z_{A_2XY}^A Z_{ABY_2}^B Z_{ABXY}^A Z_{ABXY}^X Z_{B_2XY}^B - Z_{A_2XY}^A Z_{ABY_2}^A Z_{ABXY}^B Z_{ABXY}^X Z_{B_2XY}^B + 2 Z_{A_2XY}^X Z_{ABY_2}^A Z_{ABXY}^A Z_{ABXY}^B Z_{B_2XY}^B))/(8 Z_{ABXY}^A Z_{ABXY}^B Z_{ABXY}^X (Z_{A_2XY}^A Z_{ABY_2}^B Z_{B_2XY}^X + Z_{A_2XY}^X Z_{ABY_2}^A Z_{B_2XY}^B)) + (Z_{B_2XY}^B (Z_{A_2XY}^A Z_{ABX_2}^B Z_{ABXY}^A Z_{ABXY}^Y Z_{B_2XY}^Y - Z_{A_2XY}^Y Z_{ABX_2}^A Z_{ABXY}^B Z_{ABXY}^Y Z_{B_2XY}^Y + 2 Z_{A_2XY}^Y Z_{ABX_2}^A Z_{ABXY}^A Z_{ABXY}^B Z_{B_2XY}^Y))/(8 Z_{ABXY}^A Z_{ABXY}^B Z_{ABXY}^Y (Z_{A_2XY}^A Z_{ABX_2}^B Z_{B_2XY}^Y + Z_{A_2XY}^Y Z_{ABX_2}^A Z_{B_2XY}^B)) + (Z_{B_2XY}^X (Z_{ABX_2}^A Z_{ABY_2}^Y Z_{ABXY}^A Z_{ABXY}^X Z_{B_2XY}^Y - 2 Z_{ABX_2}^A Z_{ABY_2}^A Z_{ABXY}^X Z_{ABXY}^Y Z_{B_2XY}^Y + Z_{ABY_2}^A Z_{ABX_2}^X Z_{ABXY}^A Z_{ABXY}^Y Z_{B_2XY}^Y))/(4 Z_{ABXY}^A Z_{ABXY}^X Z_{ABXY}^Y (Z_{ABX_2}^A Z_{ABY_2}^Y Z_{B_2XY}^X + Z_{ABY_2}^A Z_{ABX_2}^X Z_{B_2XY}^Y))$  (A2)

$\bar{o} = (Z_{ABY_2}^B (2 Z_{A_2XY}^X Z_{ABX_2}^B Z_{ABY_2}^Y Z_{ABXY}^X Z_{ABXY}^Y - Z_{A_2XY}^X Z_{ABX_2}^X Z_{ABY_2}^Y Z_{ABXY}^B Z_{ABXY}^Y + Z_{A_2XY}^Y Z_{ABX_2}^X Z_{ABY_2}^Y Z_{ABXY}^B Z_{ABXY}^X))/(8 Z_{ABXY}^B Z_{ABXY}^X Z_{ABXY}^Y (Z_{A_2XY}^X Z_{ABX_2}^B Z_{ABY_2}^Y + Z_{A_2XY}^Y Z_{ABY_2}^B Z_{ABX_2}^X)) + (Z_{ABY_2}^B (Z_{A_2XY}^A Z_{ABY_2}^A Z_{ABXY}^B Z_{ABXY}^X Z_{B_2XY}^X - 2 Z_{A_2XY}^X Z_{ABY_2}^A Z_{ABXY}^A Z_{ABXY}^B Z_{B_2XY}^X + Z_{A_2XY}^X Z_{ABY_2}^A Z_{ABXY}^A Z_{ABXY}^X Z_{B_2XY}^B))/(4 Z_{ABXY}^A Z_{ABXY}^B Z_{ABXY}^X (Z_{A_2XY}^A Z_{ABY_2}^B Z_{B_2XY}^X + Z_{A_2XY}^X Z_{ABY_2}^A Z_{B_2XY}^B)) + (Z_{ABY_2}^A Z_{ABY_2}^Y (2 Z_{ABX_2}^A Z_{ABXY}^X Z_{ABXY}^Y Z_{B_2XY}^X + Z_{ABX_2}^X Z_{ABXY}^A Z_{ABXY}^X Z_{B_2XY}^Y - Z_{ABX_2}^X Z_{ABXY}^A Z_{ABXY}^Y Z_{B_2XY}^X))/(8 Z_{ABXY}^A Z_{ABXY}^X Z_{ABXY}^Y (Z_{ABX_2}^A Z_{ABY_2}^Y Z_{B_2XY}^X + Z_{ABY_2}^A Z_{ABX_2}^X Z_{B_2XY}^Y))$  (A3)

$$p=(Z^X_{A_2XY}Z^Y_{A_2XY}(Z^B_{ABX_2}Z^Y_{ABY_2}Z^B_{ABXY}Z^X_{ABXY}-2Z^B_{ABX_2}Z^B_{ABY_2}Z^X_{ABXY}Z^Y_{ABXY}+Z^B_{ABY_2}Z^X_{ABX_2}$$
$$Z^B_{ABXY}Z^Y_{ABXY}))/(4Z^B_{ABXY}Z^X_{ABXY}Z^Y_{ABXY}(Z^X_{A_2XY}Z^B_{ABX_2}Z^Y_{ABY_2}+Z^Y_{A_2XY}Z^B_{ABY_2}Z^X_{ABX_2}))+$$
$$(Z^A_{A_2XY}Z^X_{A_2XY}(Z^A_{ABY_2}Z^B_{ABXY}Z^X_{ABXY}Z^B_{B_2XY}+2Z^B_{ABY_2}Z^A_{ABXY}Z^B_{ABXY}Z^X_{B_2XY}-Z^B_{ABY_2}Z^A_{ABXY}$$
$$Z^X_{ABXY}Z^B_{B_2XY}))/(8Z^A_{ABXY}Z^B_{ABXY}Z^X_{ABXY}(Z^A_{A_2XY}Z^B_{ABY_2}Z^X_{B_2XY}+Z^X_{A_2XY}Z^A_{ABY_2}Z^B_{B_2XY}))+$$
$$(Z^A_{A_2XY}Z^Y_{A_2XY}(Z^A_{ABX_2}Z^B_{ABXY}Z^Y_{ABXY}Z^B_{B_2XY}+2Z^B_{ABX_2}Z^A_{ABXY}Z^B_{ABXY}Z^Y_{B_2XY}-Z^B_{ABX_2}$$
$$Z^A_{ABXY}Z^Y_{ABXY}Z^B_{B_2XY}))/(8Z^A_{ABXY}Z^B_{ABXY}Z^Y_{ABXY}(Z^A_{A_2XY}Z^B_{ABX_2}Z^Y_{B_2XY}+Z^Y_{A_2XY}Z^A_{ABX_2}Z^B_{B_2XY})) \quad (A4)$$

All the substance quantities and their relations involved in the MQMQA are reinterpreted as follows. For ions, the expressions are only given for A and X; for pairs and quadruplets, they are only given for AX and $A_2X_2$. Readers should be familiar with how to derive similar forms for the remaining quantities. $n_A$, $n_B$, $n_X$ and $n_Y$ are the pre-input constraint conditions of composition for the ABXY solution. The mole fractions of ions, pairs and quadruplets are defined as,

$$X_A = \frac{n_A}{n_A+n_B} \quad (A5)$$

$$X_X = \frac{n_X}{n_X+n_Y} \quad (A6)$$

$$X^*_{AX} = \frac{n^*_{AX}}{\sum n^*_{ij}} \quad (A7)$$

$$X_{AX} = \frac{n_{AX}}{\sum n_{ij}} \quad (A8)$$

$$X_{A_2X_2} = \frac{n_{A_2X_2}}{\sum n_{ijkl}} \quad (A9)$$

where $\sum n_{ij}$ and $\sum n_{ijkl}$ are the total number of moles of pairs and quadruplets, respectively, equation (A7) and equation (A8) both define the FNN pair (AX) fractions but with different definitions of $\zeta$ as described below. The relation between $n_i$ and the mole quantities of the correlated quadruplets is,

$$n_A = \frac{2n_{A_2X_2}}{Z^A_{A_2X_2}} + \frac{2n_{A_2Y_2}}{Z^A_{A_2Y_2}} + \frac{2n_{A_2XY}}{Z^A_{A_2XY}} + \frac{n_{ABX_2}}{Z^A_{ABX_2}} + \frac{n_{ABY_2}}{Z^A_{ABY_2}} + \frac{n_{ABXY}}{Z^A_{ABXY}} \quad (A10)$$

$$n_X = \frac{2n_{A_2X_2}}{Z^X_{A_2X_2}} + \frac{2n_{B_2X_2}}{Z^X_{B_2X_2}} + \frac{2n_{ABX_2}}{Z^X_{ABX_2}} + \frac{n_{A_2XY}}{Z^X_{A_2XY}} + \frac{n_{B_2XY}}{Z^X_{B_2XY}} + \frac{n_{ABXY}}{Z^X_{ABXY}} \quad (A11)$$

The overall SNN coordination of A in solution is $Z_A$ which depends on the composition of the solution. Since each quadruplet only contains one SNN pair, $Z_A$ is also the number of quadruplets emanating from ion A. The following mass balance equations thus result,

$$Z_An_A = 2n_{A_2X_2} + 2n_{A_2Y_2} + 2n_{A_2XY} + n_{ABX_2} + n_{ABY_2} + n_{ABXY} \quad (A12)$$

$$Z_X n_X = 2n_{A_2X_2} + 2n_{B_2X_2} + 2n_{ABX_2} + n_{A_2XY} + n_{B_2XY} + n_{ABXY} \tag{A13}$$

$$Z_A n_A + Z_B n_B = Z_X n_X + Z_Y n_Y = 2\sum n_{ijkl} \tag{A14}$$

where equation (A14) can be derived from equations A(12-13), with similar forms for ions B and Y. The "coordination-equivalent" fractions are defined as,

$$Y_A = \frac{Z_A n_A}{Z_A n_A + Z_B n_B} = X_{A_2X_2} + X_{A_2Y_2} + X_{A_2XY} + \frac{1}{2}X_{ABX_2} + \frac{1}{2}X_{ABY_2} + \frac{1}{2}X_{ABXY} \tag{A15}$$

$$Y_X = \frac{Z_X n_X}{Z_X n_X + Z_Y n_Y} = X_{A_2X_2} + X_{B_2X_2} + X_{ABX_2} + \frac{1}{2}X_{A_2XY} + \frac{1}{2}X_{B_2XY} + \frac{1}{2}X_{ABXY} \tag{A16}$$

There are also four types of FNN pairs whose mole quantities have to be correlated with those of the quadruplets by employing the coefficient $\zeta$. $\zeta_{AX}$ is defined as the number of quadruplets emanating from an AX FNN pair. The number of moles of AX pairs could thus be expressed as

$$\zeta_{AX} n_{AX}^* = 4n_{A_2X_2} + 2n_{ABX_2} + 2n_{A_2XY} + n_{ABXY} \tag{A17}$$

where the default value of $\zeta_{AX}$ can be calculated as,

$$\zeta_{AX} = \frac{2Z_{A_2X_2}^A Z_{A_2X_2}^X}{Z_{A_2X_2}^A + Z_{A_2X_2}^X} \tag{A18}$$

As can be expected, $\zeta_{AX}$, $\zeta_{BX}$, $\zeta_{AY}$ and $\zeta_{BY}$ may have different values, but this makes the model unnecessarily complex. Therefore, a universal $\zeta$ for all types of FNN pairs has been adopted in the formalism of configurational entropy. It is defined as

$$\zeta = \frac{2Z_i}{Z_i} \tag{A19}$$

where $\zeta$ is actually defined as twice the ratio of SNN to FNN pairs, and the denominator $z_i$ represents the overall FNN coordination number of ion $i$ in the solution. Therefore, the connections between ions and their related FNN pairs can be displayed as,

$$Z_A n_A = n_{AX}^* + n_{AY}^* \tag{A20}$$

$$Z_X n_X = n_{AX}^* + n_{BX}^* \tag{A21}$$

$$F_A = \frac{Z_A n_A}{Z_A n_A + Z_B n_B} = X_{AX}^* + X_{AY}^* \tag{A22}$$

$$F_X = \frac{Z_X n_X}{Z_X n_X + Z_Y n_Y} = X_{AX}^* + X_{BX}^* \tag{A23}$$

When $\zeta$ is unified for all the ions, the following equations are generated,

$$n_{AX}^* = n_{AX} \tag{A24}$$

$$X_{AX}^* = X_{AX} = X_{A_2X_2} + \frac{1}{2}X_{ABX_2} + \frac{1}{2}X_{A_2XY} + \frac{1}{4}X_{ABXY} \tag{A25}$$

$$F_A = Y_A \qquad (A26)$$

$$F_X = Y_X \qquad (A27)$$

$$\sum n_{ij} = (4/\zeta) \sum n_{ijkl} \qquad (A28)$$

To date, we have given definitions of all the substance quantities used in the formalism of the configurational entropy. The formalism could then be expressed as just containing the internal variables of $n_{ijkl}$. All these internal variables can be determined by minimizing the Gibbs energy of the reciprocal solution under the mass constraints of A, B, X and Y.

The formalism of configurational entropy is constructed based on the distribution of all the quadruplets randomly over "quadruplet positions". Unfortunately, letting $\Delta S^{config}$ equal to $-R \sum (n_{ijkl} \ln X_{ijkl})$ would clearly overcount the number of possible configurations. Actually, there is no exact mathematical expression for $\Delta S^{config}$. Hence, approximations have to be made as equation (67) shows. There are three parts in equation (67) for $\Delta S^{config}$. The first part is regarding the ideal entropy of mixing from the point approximation, the second one regarding that from the FNN pair approximation and the third one regarding that from the quadruplet approximation. For the second part, the pair fractions $X_{ij}^*$ are divided by their values when there is no FNN SRO. A similar scheme is used for the third part, and so the quadruplet fractions $X_{ijkl}$ are also divided by their values when there is no SNN SRO. In a one-dimensional lattice, there are four distinguishable types of quadruplets: [AXBY]$_{quad}$, [AYBX]$_{quad}$, [BXAY]$_{quad}$ and [BYAX]$_{quad}$. This spawns four configurational entropy terms:

$$n_{AXBY} \ln \left( \frac{X_{AXBY}}{X_{AX} X_{BX} X_{BY}} Y_X Y_B \right) + n_{AYBX} \ln \left( \frac{X_{AYBX}}{X_{AY} X_{BY} X_{BX}} Y_Y Y_B \right) + \cdots \qquad (A29)$$

where the probability of an [AXBY]$_{quad}$ without SNN SRO is equal to $X_{AX}(X_{BX}/Y_X)(X_{BY}/Y_B)$ in which $X_{AX}$ is the probability of an AX pair, $(X_{BX}/Y_X)$ is the conditional probability of a BX pair given that one co-member of the pairs is an X, and $(X_{BY}/Y_B)$ is the conditional probability of a BY pair given that one co-member of the pairs is a B. A similar feature can be derived for the other three terms. In three dimensions, the four quadruplets are indistinguishable; hence the following equations are generated,

$$n_{AXBY} = n_{AYBX} = n_{BXAY} = n_{BYAX} = n_{ABXY}/4 \qquad (A30)$$

$$X_{AXBY} = X_{AYBX} = X_{BXAY} = X_{BYAX} = X_{ABXY}/4 \qquad (A31)$$

Substitution into equation (A29) then gives an entropy term as,

$$n_{ABXY}\ln\left(\frac{X_{ABXY}}{4X_{AX}^{\frac{3}{4}}X_{BX}^{\frac{3}{4}}X_{AY}^{\frac{3}{4}}X_{BY}^{\frac{3}{4}}}Y_A^{\frac{1}{2}}Y_B^{\frac{1}{2}}Y_X^{\frac{1}{2}}Y_Y^{\frac{1}{2}}\right) \quad (A32)$$

which is the final term within the third part of Eq.66. The other terms of the third part could be derived similarly for the cases where X=Y and/or A=B. For the sake of simplicity in these terms, the universal $\zeta$ is used and finally cancels out. Consequently, the third part of the configurational entropy (Eq.66) will become zero when there is no SNN SRO. It is noted that diverse $\zeta$, namely $\zeta_{ij}$, are used within the second part to provide more flexibility to describe the configurational entropy when strong FNN SRO exists.

## Acknowledgements

The great gratitude is owed to Emeritus Professor Arthur D. Pelton for the fruitful discussions on the recent improvements of the Modified Quasichemical Model and for his careful revision of this paper.

## References

1. Zi-Kui Liu, Computational thermodynamics and its applications, Acta Materialia, 2020, 200, 745-792.
2. William L. Bragg, Evan J. Williams, The effect of thermal agitation on atomic arrangement in alloys, Proceedings of the Royal Society of London. Series A, 1934, 145, 699-730.
3. William L. Bragg, Evan J. Williams, The Effect of Thermal Agitation on Atomic Arrangement in Alloys II, Proceedings of the Royal Society of London. Series A, 1935, 151, 540-566.
4. Ralph H. Fowler, Edward A. Guggenheim, Statistical Thermodynamics, Cambridge University Press, Cambridge, United Kingdom, 1939, 350-366.
5. Kun Wang, Xiangcheng Kong, Junlin Du, Chonghe Li, Zhilin Li, Zhu Wu, Thermodynamic description of the Ti-H system, Calphad, 2010, 34, 317-323.
6. Xiang Li, Kun Wang, Leidong Xie, Thermodynamic modeling of the GdF$_3$-MF (M: Li, K, Rb, Cs) systems, Fluid Phase Equilibria, 2017, 449, 18-27.
7. Xiang Li, Kun Wang, Mengya Xie, Zhu Wu, Leidong Xie, Thermodynamic and phase diagram modeling of CsF-MF$_4$ (M=U, Th) systems, Chemical Research in Chinese Universities, 2017, 33, 454-459.

8. Arthur D. Pelton, Sergei A. Decterov, Gunnar Eriksson, Christian Robelin, and Yves Dessureault, The Modified Quasichemical Model I—Binary Solutions, Metallurgical and Materials Transactions B, 2000, 31B, 651-659.

9. Arthur D. Pelton, Patrice Chartrand, The Modified Quasi-Chemical Model: Part II. Multicomponent Solutions, Metallurgical and Materials Transactions A, 2001, 32A, 1355-1360.

10. Patrice Chartrand, Arthur D. Pelton, The Modified Quasichemical Model: Part III. Two Sublattices, Metallurgical and Materials Transactions A, 2001, 32A, 1397-1407.

11. Arthur D. Pelton, Patrice Chartrand, Gunnar Eriksson, The modified quasichemical Model: Part IV. Two-sublattice quadruplet approximation, Metallurgical and Materials Transactions A, 2001, 32A, 1409-1416.

12. In-Ho Jung, Sergei A. Decterov, Arthur D. Pelton, Critical thermodynamic evaluation and optimization of the $MgO-Al_2O_3$, $CaO-MgO-Al_2O_3$, and $MgO-Al_2O_3-SiO_2$ Systems, Journal of Phase Equilibria and Diffusion, 2004, 25, 329-345

13. In-Ho Jung, Sergei A. Decterov, Arthur D. Pelton, Critical thermodynamic evaluation and optimization of the $CaO-MgO-SiO_2$ system, Journal of the European Ceramic Society, 2005, 25, 313-333.

14. In-Ho Jung, Sergei A. Decterov, Arthur D. Pelton, Hyun-Min Kim, Youn-Bae Kang, Thermodynamic evaluation and modeling of the Fe-Co-O system, Acta Materialia, 2004, 52, 507-519.

15. In-Ho Jung, Sergei A. Decterov, Arthur D. Pelton, Thermodynamic modeling of the $CoO-SiO_2$ and $CoO-FeO-Fe_2O_3-SiO_2$ systems, International Journal of Materials Research, 2007, 98, 816-825.

16. Patrice Chartrand, Arthur D. Pelton, Thermodynamic evaluation and optimization of the $LiCl-NaCl-KCl-RbCl-CsCl-MgCl_2-CaCl_2$ system using the modified quasi-chemical model, Metallurgical and Materials Transactions A, 2001, 32, 1361-1383.

17. Patrice Chartrand, Arthur D. Pelton, Thermodynamic evaluation and optimization of the $LiF-NaF-KF-MgF_2-CaF_2$ system using the modified quasi-chemical model, Metallurgical and Materials Transactions A, 2001, 32, 1385-1396.

18. Patrice Chartrand, Arthur D. Pelton, Thermodynamic evaluation and optimization of the Li, Na, K, Mg, Ca//F, Cl reciprocal system using the modified quasi-chemical model, Metallurgical and Materials Transactions A, 2001, 32, 1417-1430.

19. Patrice Chartrand, Arthur D. Pelton, A predictive thermodynamic model for the Al-NaF-AlF$_3$-CaF$_2$-Al$_2$O$_3$ system, Light metals-warrendale-proceedings, 2002, 245-252.

20. Kun Wang, Patrice Chartrand, A thermodynamic description for water, hydrogen fluoride and hydrogen dissolutions in cryolite-base molten salts, Phys. Chem. Chem. Phys., 2018, 20, 17324-17341.

21. Kun Wang, Zejie Fei, Jian Wang, Zhu Wu, Chonghe Li, Leidong Xie, Thermodynamic description of the AgCl-CoCl$_2$-InCl$_3$-KCl system, Materials Chemistry and Physics, 2015, 163, 73-87.

22. Kun Wang, Christian Robelin, Zhu Wu, Chonghe Li, Leidong Xie, Patrice Chartrand, Thermodynamic description of the AgCl-CoCl$_2$-InCl$_3$-NaCl system, Journal of Alloys and Compounds, 2016, 663, 885-898.

23. Kun Wang, Christian Robelin, Liling Jin, Xiaoqin Zeng, Patrice Chartrand, Thermodynamic description of the K, Be//F, Cl salt system with first-principles calculations, Journal of Molecular Liquids, 2019, 292, 111384.

24. Youn-Bae Kang, Arthur D. Pelton, Patrice Chartrand, Philip Spencer, Carlton D. Fuerst, Thermodynamic database development of the Mg-Ce-Mn-Y system for Mg alloy design, Metallurgical and Materials Transactions A, 2007, 38, 1231-1243.

25. In-Ho Jung, Zhijun Zhu, Junghwan Kim, Jian Wang, Patrice Chartrand, Arthur D. Pelton, Recent progress on the FactSage thermodynamic database for new Mg alloy development, JOM, 2017, 69, 1052-1059.

26. Zhanmin Cao, Wei Xie, Patrice Chartrand, Shenghui Wei, Guangwei Du, Zhiyu Qiao, Thermodynamic assessment of the Bi-alkali metal (Li, Na, K, Rb) systems using the modified quasichemical model for the liquid phase, Calphad, 2014, 46, 159-167.

27. Jake W. Mcmurray, Theodore M. Besmann, Johnathan Ard, Bernie FItzpatrick, Markus Piro, Jim Jerden, Mark Williamson, Benjamin S. Collins, Benjamin R. Betzler, Arthur L Qualls, Multi-physics simulations for molten salt reactor evaluation: Chemistry modeling and database development, Oak Ridge National Laboratory, ORNL/SPR-2018/864, Oak Ridge, TN (2018).

28. Arthur D. Pelton, A general "geometric" thermodynamic model for multicomponent solutions, Calphad, 2001, 25, 319-328.
29. Patrice Chartrand, Arthur D. Pelton, On the choice of "geometric" thermodynamic models, Journal of Phase Equilibria, 2000, 21, 141-147.
30. Guillaume Lambotte, Patrice Chartrand, Thermodynamic optimization of the ($Na_2O+SiO_2+NaF+SiF_4$) reciprocal system using the Modified Quasichemical Model in the Quadruplet Approximation, Journal of Chemical Thermodynamics, 2011, 43, 1678-1699.
31. Arthur D. Pelton, Phase diagrams and thermodynamic modeling of solutions, Elsevier, 2019.